\documentclass[a4paper,11pt]{article}
\usepackage{jheppub}

\usepackage[utf8]{inputenc}
\usepackage{graphicx,cancel,float}
\usepackage{amssymb}
\usepackage{textcomp}
\usepackage{amsmath}
\usepackage{tabularx}
\usepackage{bm}
\usepackage{times}
\usepackage{color}
\usepackage{slashed}
\usepackage{multirow}
\usepackage{verbatim}
\usepackage{epsfig,epstopdf}
\usepackage{subfigure}

\allowdisplaybreaks[4]

\def\lsim{\mathrel{\raise.3ex\hbox{$<$\kern-.75em\lower1ex\hbox{$\sim$}}}}
\def\gsim{\mathrel{\raise.3ex\hbox{$>$\kern-.75em\lower1ex\hbox{$\sim$}}}}

\newcommand{\braket}[1]{\ensuremath{\left\langle #1 \right\rangle}}

\definecolor{orange}{rgb}{1,0.5,0}

\preprint{}
\subheader{\it \today}

\title{Probing general $U(1)'$ models with non-universal lepton charges at FASER/FASER2, COHERENT and long-baseline oscillation experiments}

\author[a]{Tobias Felkl}
\emailAdd{t.felkl@unsw.edu.au}
\affiliation[a]{
Sydney Consortium for Particle Physics and Cosmology,\\
School of Physics, The University of New South Wales, Sydney, New South Wales 2052, Australia
}

\author[b]{Tong Li}
\emailAdd{litong@nankai.edu.cn}
\affiliation[b]{
School of Physics, Nankai University, Tianjin 300071, China
}
\author[c]{Jiajun Liao}
\emailAdd{liaojiajun@mail.sysu.edu.cn}
\affiliation[c]{
School of Physics, Sun Yat-Sen University, Guangzhou 510275, China
}
\author[a]{Michael A. Schmidt}
\emailAdd{m.schmidt@unsw.edu.au}
\preprint{CPPC-2023-03}
\abstract{
The general anomaly-free $U(1)'$ models allow non-universal lepton charges.
We explore the sensitivities of FASER/FASER2, COHERENT and DUNE/T2HK  precision experiments to the new gauge boson $Z'$ and the new CP-even scalar $\phi$.
With non-universal lepton charges, distinctive reaches at FASER/FASER2 emerge in the regime of low $m_{Z'}$ and small gauge coupling $g_{BL}$ for different $U(1)'$ charge setups. The COHERENT experiment and the future long-baseline experiments DUNE/T2HK also provide complementary probes to the available parameter space.
For $m_\phi < 2m_{Z'}$, the search for the scalar $\phi$ at FASER/FASER2 is sensitive to the mixing angle between the scalar singlet and the SM Higgs.
In the case of $m_\phi > 2m_{Z'}$, the kinematically allowed decay $\phi\to Z' Z'$ changes the lifetime and decay rates of the scalar $\phi$. The sensitivity reach highly depends on the $Z'$ mass and the gauge coupling $g_{BL}$.
}

\arxivnumber{arXiv:2306.09569}

\begin{document}

\maketitle
\setcounter{page}{2}

\newpage

\section{Introduction}
\label{sec:Intro}

Neutrino oscillations have been observed by various neutrino experiments
in the last two decades. Their explanation requires non-vanishing neutrino masses which cannot be accounted for by the Standard Model (SM).
It thus provides a strong motivation for new physics beyond the SM associated with neutrinos.
The most economical mechanism for generating neutrino masses is the seesaw mechanism~\cite{Minkowski:1977sc,Yanagida:1979as,GellMann:1980vs,Glashow:1979nm,Mohapatra:1979ia,Shrock:1980ct,Schechter:1980gr}, in which one
introduces at least two right-handed neutrinos. Right-handed neutrinos also allow the introduction of an extra Abelian gauge symmetry $U(1)'$ by choosing their charges to cancel gauge anomalies~\cite{Carlson:1986cu}.
The simplest and best-studied anomaly-free formulation for a gauged $U(1)'$ would be the $U(1)_{B-L}$ extension~\cite{Davidson:1978pm,Mohapatra:1980qe,Marshak:1979fm,Wetterich:1981bx,Masiero:1982fi,Mohapatra:1982xz,Buchmuller:1991ce}, where $B$ and $L$ stand
for baryon number and lepton number, respectively. In this model, the $U(1)'$ has no $U(1)_Y$ component and the charges are the same for three generations of SM fermions, i.e., $B=1/3$ for SM quarks and $L=-1$ for charged leptons and neutrinos.

This is however only one special case. The anomaly conditions allow non-universal $U(1)'$ charges for different generations of the SM leptons.
The non-universal gauge couplings lead to a non-trivial flavor structure for the fermion masses and may in particular lead to texture zeros in the neutrino Yukawa couplings and the Majorana mass matrix~\cite{Araki:2012ip,Liao:2013rca,Bhatia:2017tgo,Bonilla:2017lsq,Babu:2017olk,Ellis:2017nrp,Alonso:2017uky,Bhatia:2021eco}. Phenomenologically, the general $U(1)'$ models with non-universal lepton charges can result in distinguishable searches in high-intensity experiments and may evade the severe constraints by electron beam-dump experiments~\cite{Abdallah:2020biq,Bauer:2020itv,Kling:2020iar,delaVega:2021wpx,Coy:2021wfs,Amaral:2021rzw}. Moreover, the new scalar field spontaneously breaking the $U(1)'$ symmetry is not necessarily very heavy and can be naturally below 1 TeV or even much lighter~\cite{Iso:2009ss,Okada:2012sg,Dev:2021qjj}. The new light scalar in the same model can also be probed through the mixing with the SM Higgs boson.

The newly predicted particles in $U(1)'$ models with very weak couplings would become long-lived and can be searched for at the Forward Physics Facility (FPF) such as the far-forward experiments at the Large Hadron Collider (LHC). FASER (ForwArd Search ExpeRiment) is an on-going experiment in the far-forward region of ATLAS and located about 480 m from the ATLAS interaction point~\cite{Feng:2017uoz,FASER:2018eoc}. The detector will be upgraded to FASER2 in the high luminosity LHC (HL-LHC) era. The long-lived particles (LLPs) can be produced by proton-beam bremsstrahlung and the rare decays of mesons, and then decay in the detector. The searches for LLPs such as a dark photon, a dark Higgs as well as the vector boson in $U(1)_{B-L}$ model have been simulated at FASER/FASER2~\cite{Feng:2017vli,FASER:2018eoc,Boiarska:2019vid,Araki:2020wkq,Li:2021rzt,Asai:2022zxw}. We expect that the new gauge boson in $U(1)'$ models with non-universal gauge couplings would lead to distinctive search reaches.

The $U(1)'$ models also introduce non-standard interactions (NSIs) between neutrinos and other SM fermions via neutral current interactions after integrating out the new gauge boson.
Coherent elastic neutrino-nucleus scattering (CE$\nu$NS)~\cite{PhysRevD.9.1389} and precision neutrino oscillation experiments can probe the new physics in such NSIs. The CE$\nu$NS process was first observed by the COHERENT experiment in the Spallation Neutron Source (SNS) at the Oak Ridge National Laboratory~\cite{COHERENT:2017ipa}. The neutrinos measured at COHERENT are produced by the decays of stopped pions
and muons through $\pi^+\to \mu^+\nu_\mu$ and $\mu^+\to e^+\nu_e \bar{\nu}_\mu$. Thus, the COHERENT results would be sensitive to the $U(1)'$ models with gauge couplings to electron or muon neutrinos~\cite{Heeck:2018nzc,Flores:2020lji,Cadeddu:2020nbr,delaVega:2021wpx,Cheung:2021tmx,AtzoriCorona:2022moj}. Moreover, due to the change of neutrino propagation in matter in terms of neutral current NSIs, the long-baseline neutrino oscillation experiments such as DUNE are also sensitive to probe the NSIs in the $U(1)'$ models~\cite{Flores:2020lji,Chatterjee:2021wac,Asai:2022zxw}.

In this work, we study general $U(1)'$ models with non-universal lepton charges. A new scalar singlet is introduced with non-zero $U(1)'$ charge. After the scalar acquires a vacuum expectation value, the $U(1)'$ symmetry is broken and the masses of the neutral gauge boson $Z'$ and of Majorana neutrinos are generated. The scalar is mixed with the SM Higgs doublet and there are two CP-even physical Higgses, i.e., the SM-like $h$ and $\phi$.
We then consider the sensitivities of above precision experiments to the new particles in three scenarios~\cite{Dev:2021qjj}:
\begin{itemize}
\item $Z'$ production and decay with non-universal lepton charges\;;
\item $\phi$ production and decay for $m_\phi < 2m_{Z'}$ or $m_\phi > 2m_{Z'}$ with $\phi\to Z' Z'$ open\;;
\item $Z'$ production when $m_\phi > 2m_{Z'}$ with $\phi\to Z' Z'$ open\;.
\end{itemize}
In the first scenario, the non-universal lepton charges affect the lifetime and decay rates of the gauge boson $Z'$. We expect distinctive reaches in the regime of low $m_{Z'}$ and small gauge couplings for different $U(1)'$ charge setups. The dimensionless coupling
parameters in the NSIs are scaled as $\epsilon\sim g^2_{BL}v^2/m_{Z'}^2$ where $v\approx 246$ GeV is the electroweak vacuum expectation value and $g_{BL}$ is the $U(1)'$ gauge coupling. The COHERENT experiment and the neutrino oscillation experiments also provide complementary reach with a linear dependence on $g_{BL}^2/m_{Z'}^2$.
In the second scenario with $m_\phi < 2m_{Z'}$, analogous to the dark Higgs case~\cite{Feng:2017vli}, the search reach of $\phi$ is sensitive to the mixing angle with the SM Higgs boson. However, if $m_\phi > 2m_{Z'}$, the kinematically allowed decay $\phi\to Z' Z'$ changes the lifetime and decay rates of the scalar $\phi$. The probe of $\phi$ depends on additional parameters which are the $Z'$ mass and the gauge coupling $g_{BL}$. Moreover, with $\phi\to Z' Z'$ open in the third scenario, additional production modes of the $Z'$ appear.

This paper is organized as follows. In Sec.~\ref{sec:U1} we introduce the general $U(1)'$ models and discuss their details which are essential for the phenomenological studies below.
In Sec.~\ref{sec:FASERZp} we study the sensitivities of FASER/FASER2 and neutrino experiments to a light $Z'$ in general $U(1)'$ models with non-universal lepton charges. Then, we investigate the search for $\phi$ and the improved $Z'$ search at FASER/FASER2 in Sec.~\ref{sec:FASERphi}.
Finally, in Sec.~\ref{sec:Con} we draw our main conclusions.

\section{The general \texorpdfstring{$U(1)'$}{U(1)'} models with non-universal leptonic charges}
\label{sec:U1}

\begin{table}
\begin{center}
\begin{tabular}{c|cccc}
Particles & $SU(3)_C$ & $SU(2)_L$ & $U(1)_Y$ & $ U(1)^\prime$\\\hline
$L_{iL}$ & 1 & 2 & $-1/2$ & $Q_i^\prime$\\
$e_{iR}$ & 1 & 1 & $-1$ & $Q_i^\prime$\\
$\nu_{iR}$ & 1 & 1 & $-1$ & $Q_{\nu i}^\prime$\\\hline
$Q_{iL}$ & 3 & 2 & $1/6$ & $1/3$ \\
$u_{iR}$ & 3 & 1 & $2/3$ & $1/3$ \\
$d_{iR}$ & 3 & 1 & $-1/3$ & $1/3$ \\ \hline
$H$ & 1 & 2 & $1/2$ & $0$ \\
$\Phi_m$ & 1 & 1 & 0 & $Q_{\Phi,m}^\prime$ \\\hline
\end{tabular}
\end{center}
\caption{Particles content and the charges of the model, where $i$
is the fermion generation index,
$H$ is the SM Higgs doublet and $\Phi_m$ denotes the additional scalar singlets.
}
\label{tab:charges}
\end{table}

We consider general $U(1)'$ models based on non-universal leptonic charges as given in Table~\ref{tab:charges}. We allow the charges of the right-handed (RH) neutrinos to be different from that of the SM leptons in contrast to Ref.~\cite{Kownacki:2016pmx} in which they were assumed to be equal. The universal quark charges of $1/3$ are taken to avoid the constraints from quark flavor-changing neutral current processes. References~\cite{Greljo:2021npi,Greljo:2022dwn} also considered non-universal lepton charges to explain flavour anomalies involving muons.
As also shown in Ref.~\cite{Kownacki:2016pmx}, the $[SU(3)_c]^2 U(1)^\prime$ and $U(1)_Y [U(1)^\prime]^2$ anomaly conditions and the usual SM anomaly conditions are automatically satisfied. The other anomaly conditions amount to
\begin{align}
[U(1)^\prime]^3 & : \sum_i \left( 2[Q_i^\prime]^3 - [Q_i^\prime]^3\right)  - \sum_i [Q_{\nu i}^\prime]^3\;,
\\
[SU(2)_L]^2 U(1)^\prime & : \frac12 \left( 3 + \sum_i Q_i^\prime\right)\;,
\\
[U(1)_Y]^2 U(1)^\prime & :  3 \left[6\left(\frac16\right)^2 -3 \left(\frac23\right)^2 -3 \left(-\frac13\right)^2\right]\frac13 + \left[2\left(-\frac12\right)^2 - (-1)^2\right] \sum_i Q_i^\prime\;,
\\
[\mathrm{gravity}]^2 U(1)^\prime &: \sum_i (2 Q_i^\prime -Q_i^\prime) -\sum_i Q_{\nu i}\;,
\end{align}
where we have taken into account the fact that the SM quarks do not contribute to the $[U(1)^\prime]^3$ and the $[\mathrm{gravity}]^2 U(1)^\prime$ anomalies due to their vectorial couplings.
The above anomaly conditions can be resolved with
\begin{align}\label{charge}
\sum_i Q_i^\prime &= -3 \;, &
\sum_i Q'_{\nu i} &=\sum_i Q_i^\prime\;, &
\sum_i Q_{\nu i}^{\prime3}&=\sum_i Q_i^{\prime3}\;.
\end{align}
As a result, there are only three independent leptonic charges and at least one of them is from the RH neutrinos.
Without loss of generality, we utilize the first two equalities to determine the charges of $Q_\tau^\prime$ and $Q_{\nu 3}^\prime$ which are $Q_\tau^\prime = -3-Q_e^\prime - Q_\mu^\prime$ and $Q_{\nu 3}^\prime = -3 -Q_{\nu 1}^\prime- Q_{\nu 2}^\prime$.
We further use the last equality to eliminate one of the RH neutrino charges $Q_{\nu 1}^\prime$ and $Q_{\nu 2}^\prime$. Thus, the free leptonic charges include $Q_e^\prime$, $Q_\mu^\prime$ as well as one independent charge for the RH neutrino.
Note that the charges do not have to be quantized and can take arbitrary real values.

The kinetic terms for the Abelian sector are given by
\begin{align}
\mathcal{L}_{gauge} & = -\frac{1}{4} F^{\mu\nu}F_{\mu\nu}-\frac{1}{ 4}F^{\prime\mu\nu}F^\prime_{\mu\nu}-\frac{\epsilon}{2}F^{\mu\nu}F^\prime_{\mu\nu}\;,
\end{align}
with $F^{\mu\nu}=\partial^\mu B^\nu - \partial^\nu B^\mu$ and $F^{\prime\mu\nu}=\partial^\mu B^{\prime\nu} - \partial^\nu B^{\prime\mu}$ being the field strength tensors. Here $B^\mu$ and $B^{\prime\mu}$ are the gauge fields for $U(1)_Y$ and $U(1)'$, respectively. Although the kinetic mixing term can be generated via charged fermion loops, it is generally negligible compared to direct couplings. Since the mixing between the Abelian gauge bosons has to be very small, we work in the case with $\epsilon = 0$. The kinetic terms for the matter fields are
\begin{eqnarray}
\mathcal{L}_{kinetic} & =& i\overline{Q}_L \gamma^\mu D_\mu Q_L + i\overline{u}_R \gamma^\mu D_\mu u_R + i\overline{d}_R \gamma^\mu D_\mu d_R \nonumber \\
&&{}+i\overline{L}_L \gamma^\mu D_\mu L_L +i\overline{e}_R \gamma^\mu D_\mu e_R +i\overline{\nu}_R \gamma^\mu D_\mu \nu_R\;,
\end{eqnarray}
where the covariant derivative for the $U(1)'$ is
\begin{eqnarray}
D_\mu f_{L,R}=\partial_\mu f_{L,R}+ig_{BL}Q'_f B'_\mu f_{L,R}\;,
\end{eqnarray}
with $f$ being the SM fermions.
The $Z^\prime$ gauge boson directly couples to quarks and leptons proportional to their charges $Q^\prime_f$ which enter in the covariant derivative $D_\mu =\partial_\mu +ig_{BL}Q'_f B'_\mu$.
In the following analysis we choose the basis where the charged lepton and up-type quark mass matrices are diagonal.

The most general renormalizable Lagrangian for the neutrino sector reads
\begin{eqnarray}
-\mathcal{L}_{Y}^\nu
& =& y_{ij}^D \overline{l_i}_L \tilde{H} \nu_{Rj} + \frac12 M_{ij} \overline{\nu_{Ri}^C} \nu_{Rj} + \frac12 y_{ij,m}^M \overline{\nu_{Ri}^C}\nu_{Rj} \Phi_m +h.c.\;,
\end{eqnarray}
where $\tilde{H}=i\sigma_2 H^\ast$ and the scalar singlet fields $\Phi_m$ carry $U(1)'$ charges $Q_{\Phi,m}^\prime$.
The Dirac and RH Majorana mass matrices are
\begin{align}
	M_D & = \frac{y^D v_0}{\sqrt{2}}\;, &
	M_R & = M + \frac{y^M_m v_{m}}{\sqrt{2}} \;,
\end{align}
where the scalar fields get their vacuum expectation values as $\braket{H}^T = (0, v_0/\sqrt{2})$ and $\braket{\Phi_m}= v_m/\sqrt{2}$.
The light neutrino mass matrix is then given by the usual seesaw expression
\begin{eqnarray}
m_\nu=-M_D M_R^{-1} M_D^T\;.
\end{eqnarray}
The observation of two non-zero neutrino masses further restricts the leptonic charges. The Dirac neutrino mass matrix $M_D$ and the RH neutrino Majorana mass matrix $M_R$ are forced to have at least rank $2$.
The condition on the rank of the Dirac neutrino mass matrix enforces $Q_{\nu i}^\prime = Q_i^\prime$ without loss of generality, taking into account the fact that the RH neutrinos can always be reordered.
Thus, this condition further removes the degree of freedom of the RH neutrino charge and there are at most two independent charges $Q_{e,\mu}^\prime$.
For the RH neutrino mass matrix, a rank 2 matrix requires at least two independent non-zero entries.
As the charges of the scalars $Q_{\Phi,m}^\prime$ are not constrained, it is always possible to generate a contribution from the Yukawa interaction by choosing $Q_{\Phi,m}^\prime$ appropriately without imposing additional constraints on the charge assignment.
For the direct mass term, the condition $Q_e^\prime=0$ ($Q_\mu^\prime=0$) [$Q_e^\prime + Q_\mu^\prime=-3$] leads to a non-zero $(1,1)$ ($(2,2)$) [$(3,3)$] entry. Similarly, the conditions $Q_e^\prime+Q_\mu^\prime=0$ ($Q_e^\prime=-3$) [$Q_\mu^\prime=-3$] individually induce non-zero off-diagonal terms in the $(1,2)$ ($(2,3)$) [$(1,3)$] entry.

For illustration, we take a model based on $U(1)_{B-3 L_e + Q^\prime L_\mu -Q^\prime L_\tau}$ with $Q^\prime\neq 3$ as well as one
scalar singlet $\Phi$ with charge $Q^\prime_\Phi$ and $\braket{\Phi}= v_1/\sqrt{2}$. This charge assignment leads to a diagonal neutrino Yukawa matrix $y^D$ with arbitrary entries. The combination of the Majorana mass matrix and the Yukawa couplings matrix $y^M$ for $Q_{\Phi}^\prime=-2 Q^\prime_e$ results in an effective Majorana mass matrix~\footnote{Although two non-zero Majorana mass entries with a diagonal Dirac neutrino mass matrix can explain neutrino masses, the observed leptonic mixing requires more non-zero entries. It can be straightforwardly generated by introducing multiple scalars $\Phi_m$. See Appendix~\ref{sec:Mixing} for a detailed discussion of neutrino Yukawa textures.}
\begin{align}
M + \frac{v_1}{\sqrt{2}} y^M  & =
\begin{pmatrix}
		\frac{v_1}{\sqrt{2}} y^{M}_{11}   & 0 & 0 \\
		.  & 0 & M_{23} \\
		.  & . & 0 \\
\end{pmatrix}
\qquad\Rightarrow\qquad
m_\nu  = - \frac{v_0^2}{2}
\begin{pmatrix}
		\frac{(y^{D}_{11})^2\sqrt{2}}{y^{M}_{11} v_1}   & 0 & 0 \\
		.  & 0 & \frac{y^D_{22} y^D_{33} }{M_{23}} \\
		.  & . & 0 \\
\end{pmatrix} \;.
\end{align}
This texture agrees with $L_\mu-L_\tau$ symmetry. Note that the charge $Q^\prime$ is not quantized and can be arbitrarily chosen. It is straightforward to construct similar models with a free charge which is not quantized. Motivated by the possibility to have free charges, in the next section we will investigate the phenomenology of a gauged $U(1)'$ symmetry with arbitrary charges $Q_{e,\mu}^\prime$ in a simple model with one scalar $\Phi$.
Even if there is no free charge, the analysis in the next section can also be generally used for models with fixed charges. For instance, the minimal $U(1)_{B-L}$ model~\cite{Carlson:1986cu} with universal $U(1)'$ charges, the $U(1)_{B-3L_e-L_\mu+L_\tau}$, $U(1)_{B-L_e-3L_\mu+L_\tau}$, or $U(1)_{B-3L_\mu}$ model are motivated
by the introduction of vanishing textures in the neutrino mass matrix~\cite{Araki:2012ip,Liao:2013rca,Bhatia:2017tgo,Bhatia:2021eco} and can have implications for non-standard neutrino interactions~\cite{Han:2019zkz,Bauer:2020itv}.

The general Lagrangian for the electroweak Higgs doublet $H$ and the scalar field $\Phi$ with $U(1)^\prime$ charge $Q_{\Phi}^\prime$ is given by
\begin{eqnarray}
\mathcal{L}_{scalar} & =& (D_\mu H)^\dagger (D^\mu H)+(D_\mu \Phi)^\dagger (D^\mu \Phi)- V(H,\Phi)\;.
\end{eqnarray}
Once getting the vacuum expectation value $\langle\Phi\rangle\to v_1/\sqrt{2}$, the extra scalar breaks the new $U(1)'$ gauge symmetry with $D_\mu \Phi=\partial_\mu \Phi+ig_{BL}Q'_\Phi B'_\mu \Phi$.
The phenomenologically relevant interactions of the scalar $\Phi$ are governed by the scalar potential
\begin{eqnarray}
V(H,\Phi)&=&m_H^2 H^\dagger H + \lambda_H (H^\dagger H)^2 +m_\Phi^2 \Phi^\dagger \Phi + \lambda_\Phi (\Phi^\dagger \Phi)^2  + \lambda_1 (H^\dagger H) (\Phi^\dagger \Phi)
\end{eqnarray}
which is parameterized by five real parameters, $m_H^2,\lambda_H, m_\Phi^2, \lambda_\Phi$ and the Higgs portal coupling $\lambda_1$. The electroweak symmetry and the $U(1)'$ symmetry are spontaneously broken.
We parameterize the SM Higgs doublet $H$ and the singlet $\Phi$ as
$H^T=(H^+,H^0)=(H^+,(v_0+h^0+iG^0)/\sqrt{2})$
and $\Phi=(v_1+\phi_1+i\chi_1)/\sqrt{2}$
with the vacuum expectation values
\begin{align}
	v_0^2 & = \frac{4m_H^2 \lambda_\Phi -2\lambda_1 m_\Phi^2}{\lambda_1^2-4\lambda_H \lambda_\Phi}\;, &
	v_1^2 & = \frac{4m_\Phi^2 \lambda_H -2\lambda_1 m_H^2}{\lambda_1^2-4\lambda_H \lambda_\Phi}\;.
\end{align}
After the spontaneous symmetry breaking, the Goldstone boson $\chi_1$ is eaten by the neutral gauge boson $Z^\prime$ and the $Z^\prime$ acquires a mass  $m_{Z'}=g_{BL} Q_\Phi^{\prime} v_1$.
The masses of the two CP-even scalars are described by
\begin{eqnarray}
\mathcal{M}^2=\begin{pmatrix}
			2\lambda_H v_0^2  & \lambda_1 v_0 v_1 \\
			\lambda_1 v_0 v_1  & 2\lambda_\Phi v_1^2 \\
		\end{pmatrix}\;.
\end{eqnarray}
The physical Higgs mass eigenstates are defined as
\begin{eqnarray}
\begin{pmatrix}
			h  \\
			\phi  \\
		\end{pmatrix}=
\begin{pmatrix}
			\cos\theta  & \sin\theta \\
			-\sin\theta  & \cos\theta \\
		\end{pmatrix}
\begin{pmatrix}
			h^0  \\
			\phi_1  \\
		\end{pmatrix}
		\qquad
		\text{with}\qquad
		\tan2\theta = \frac{\lambda_1 v_0 v_1}{\lambda_H v_0^2 - \lambda_\Phi v_1^2}\;.
\end{eqnarray}
We thus parameterize the new physics in terms of the $Z^\prime$ mass $m_{Z^\prime}$, the gauge coupling $g_{BL}$, the two CP-even scalar masses $m_h$ and $m_\phi$, and the mixing angle $\theta$. The three dimensionless parameters are
\begin{eqnarray}
\lambda_H&=& (c_\theta^2 m_h^2+s_\theta^2 m_\phi^2)/(2v_0^2)\;,\nonumber\\
\lambda_\Phi&=& (s_\theta^2 m_h^2+c_\theta^2 m_\phi^2)/(2v_1^2)\;,\nonumber\\
\lambda_1&=& c_\theta s_\theta (m_h^2 - m_\phi^2)/(v_0 v_1)\;,
\end{eqnarray}
where $s_\theta=\sin\theta$ and $c_\theta=\cos\theta$.
The two vacuum expectation values $v_0$ and $v_1$ are given in terms of the Fermi constant $v_0=(\sqrt{2} G_F)^{-1/2}$ and the ratio $v_1=m_{Z^\prime}/(g_{BL} Q_\Phi^\prime)$. If $m_\phi<m_h/2$, the SM-like Higgs boson can decay into a pair of light CP-even scalars $\phi$. The scalar $\phi$ couples to the SM particle through the small mixing angle $\theta$ and can be long-lived. The Higgs invisible width is given by
\begin{eqnarray}
\label{eq:Hphiphi}
\Gamma(h\to \phi\phi) = \frac{\lambda_{h\phi\phi}^2}{8\pi m_h} \sqrt{1-\frac{4m_\phi^2}{m_h^2}}\;,\qquad
\lambda_{h\phi\phi} = \frac{s_{2\theta}}{4 v_0 v_1}(m_h^2+2m_\phi^2)(v_0 c_\theta + v_1 s_\theta)\;.
\end{eqnarray}
The constraint on $\lambda_{h\phi\phi}$ comes from the invisible decay of Higgs boson~\cite{CMS:2018yfx,ATLAS:2019cid,ATLAS:2021gcn,CMS:2022qva} and the latest limit on the corresponding decay branching fraction is 0.18 at the 95\% CL~\cite{CMS:2022qva}.

As outlined above in the phenomenological analysis we focus on the $U(1)'$ models which consist of three right-handed neutrinos, the $Z^\prime$ gauge boson and a (light) scalar field $\Phi$ which breaks the $U(1)'$ gauge symmetry spontaneously. While we assume universal charges for the quarks $Q_{1,2,3}^\prime=1/3$, we employ non-universal charges $Q_i^\prime$ for the lepton fields $L_{Li}$, $e_{Ri}$ and $\nu_{Ri}$.
The kinetic mixing is neglected in the analysis because its contribution is negligible, and thus the couplings of the $Z^\prime$ gauge boson are entirely fixed by the charges of the SM fermions.
We also assume decoupled right-handed neutrinos for simplicity, so the gauge boson $Z'$ does not decay into them.
An additional partial decay width $\Gamma(\phi\to \nu_R\nu_R)$ decreases the lifetime $\tau_\phi$ and leads to a suppression of the visible branching ratio by a factor $\Gamma_{\rm visible}/(\Gamma(\phi\to \mathrm{SM})+\Gamma(\phi\to \nu_R\nu_R))$.
See Refs.~\cite{FASER:2018eoc,Berryman:2019dme,Boiarska:2021yho} for recent works on the search for heavy neutral leptons at FASER, DUNE or CHARM.

\section{Sensitivity of FASER and neutrino experiments to \texorpdfstring{$Z'$}{Z'} in general \texorpdfstring{$U(1)'$}{U(1)'} models}
\label{sec:FASERZp}

\subsection{The sensitivity of FASER/FASER2 to the gauge boson \texorpdfstring{$Z'$}{Z'}}

For the gauge boson $Z'$, as seen in the last section, both of its mass and its couplings to the SM matter particles are determined by the gauge coupling $g_{BL}$. A very weak coupling $g_{BL}$ can induce a light and long-lived $Z'$ which can be searched for in far-forward experiments such as FASER/FASER2.
Throughout this section, we assume that the scalar can be neglected, either due to its large mass or tiny scalar mixing. We will estimate and comment on the possibly additional $Z'$ contribution from $\phi\to Z' Z'$ in Sec.~\ref{sec:FASERimprovedZp}.

The major production mechanisms of $Z'$ bosons at FASER include proton-beam bremsstrahlung as well as the decays of light mesons~\cite{Feng:2017uoz}.
The $Z'$ production cross sections and meson decay widths can be easily obtained
from those of the dark photon $A'$ with $\varepsilon$ being the $\gamma-A'$ kinetic mixing, as presented in the darkcast package~\cite{Ilten:2018crw}.
The approximate ratio of the bremsstrahlung cross sections is
\begin{eqnarray}
\frac{\sigma(pZ\to pZZ')}{\sigma(pZ\to pZA')}\approx \frac{g_{BL}^2(2Q'_u+Q'_d)^2}{(\varepsilon e)^2}\;,
\end{eqnarray}
where $Q'_u=Q'_d=1/3$ in our setup.
The general decay width ratios of $\pi^0$ and $\eta$ are then given by
\begin{eqnarray}
{\Gamma(\pi^0\to Z'\gamma)\over \Gamma(\pi^0\to A'\gamma)}&=&\Big({g_{BL}\over \varepsilon e}\Big)^2 {|(Q'_u-Q'_d){\rm BW}_\rho+3(Q'_u+Q'_d){\rm BW}_\omega|^2\over |{\rm BW}_\rho+{\rm BW}_\omega|^2}\;,\\
{\Gamma(\eta\to Z'\gamma)\over \Gamma(\eta\to A'\gamma)}&=&\Big({g_{BL}\over \varepsilon e}\Big)^2 {|9(Q'_u-Q'_d){\rm BW}_\rho+3(Q'_u+Q'_d){\rm BW}_\omega+6Q'_s{\rm BW}_\phi|^2\over |9{\rm BW}_\rho+{\rm BW}_\omega-2{\rm BW}_\phi|^2}\;,
\end{eqnarray}
where ${\rm BW}_V$ denotes the Breit-Wigner form factors~\cite{Tulin:2014tya,Ilten:2018crw} for vector mesons $V=\rho,\omega,\phi$ in the framework of vector meson dominance (VMD)~\cite{Fujiwara:1984mp}. In FASER detectors, the gauge boson can decay into all kinematically accessible states. The $Z'$ with a mass smaller than $2m_\tau$ mainly decays into neutrinos, $e^+e^-$, $\mu^+\mu^-$ and light hadrons.
The partial width of the $Z'$ decay into SM final states is given by
\begin{eqnarray}
\Gamma(Z'\to f\bar{f})={C_f (g_{BL} Q'_f)^2\over 12\pi}m_{Z'}\left(1+{2m_f^2\over m_{Z'}^2}\right)\sqrt{1-{4m_f^2\over m_{Z'}^2}}\;,
\end{eqnarray}
where the color factor is $C_f=1$ for charged leptons, 3 for quarks, and $1/2$ for active neutrinos. Here we assume the right-handed neutrinos are heavy enough to decouple. The VMD approach was also used to estimate the decay width for specific hadronic final states as stated in Ref.~\cite{Ilten:2018crw}.

\begin{figure}[tb!]
\begin{center}
\includegraphics[scale=1,width=0.48\linewidth]{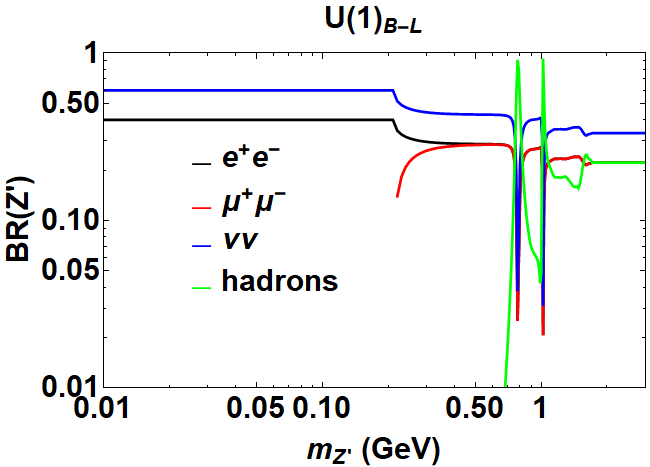}
\includegraphics[scale=1,width=0.48\linewidth]{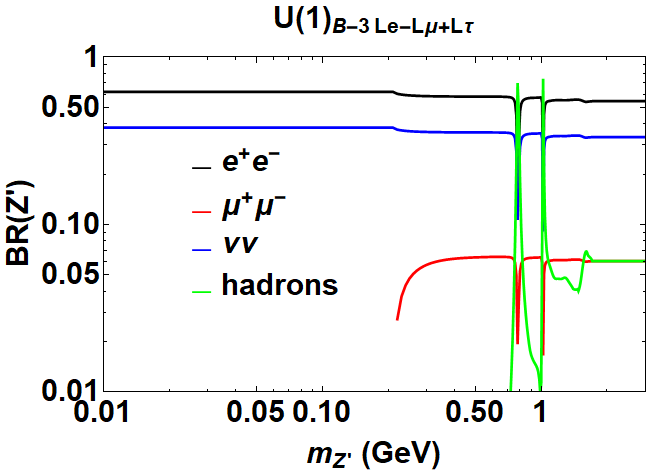}
\end{center}
\caption{The decay branching ratios of $Z'$ bosons for $U(1)_{B-L}$ (left) and $U(1)_{B-3L_e-L_\mu+L_\tau}$ (right) models.
}
\label{fig:brZp}
\end{figure}

We use the darkcast package~\cite{Ilten:2018crw} to calculate the lifetime and branching fractions of the $Z'$ in general $U(1)'$ models as shown in Fig.~\ref{fig:brZp} for illustration. Apart from the two peaks from vector meson resonances, where hadronic final states dominate, the $Z'$ boson dominantly decays to leptons.
One can see that the different charge $Q'_e$ would change the dominant decay channel and the branching fractions of viable decay products significantly. These results are then passed to the FORESEE package~\cite{Kling:2021fwx}. In the FORESEE package, we modify the production rates of the dark photon to obtain the forward $Z'$ flux and specify the allowed channels of $Z'$ decays to be those into $e^+e^-$, $\mu^+\mu^-$ and hadrons. FORESEE then provides the number of signal events passing the selection criteria in FASER detectors.
The sensitivity is then estimated by requiring 2.3 signal events to accommodate 90\% C.L. interval for the Poisson signal mean, given zero observed total events and zero mean background.
The projected gauge boson sensitivity reaches are derived using LHC Run 3 with 150 fb$^{-1}$ for FASER, and HL-LHC with 3 ab$^{-1}$ for FASER2.

\begin{figure}[tbp!]
\begin{center}
\includegraphics[scale=1,width=0.45\linewidth]{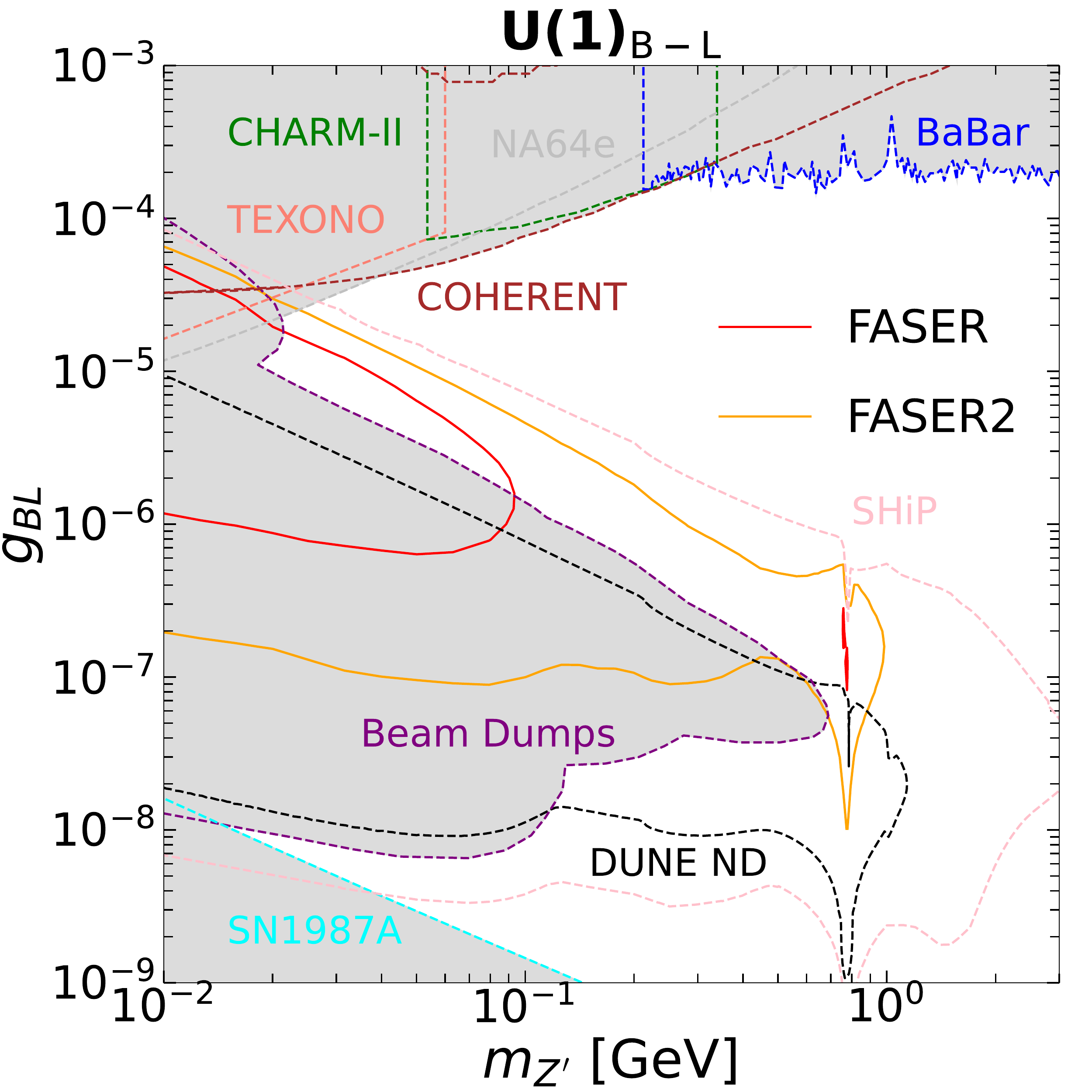}
\includegraphics[scale=1,width=0.45\linewidth]{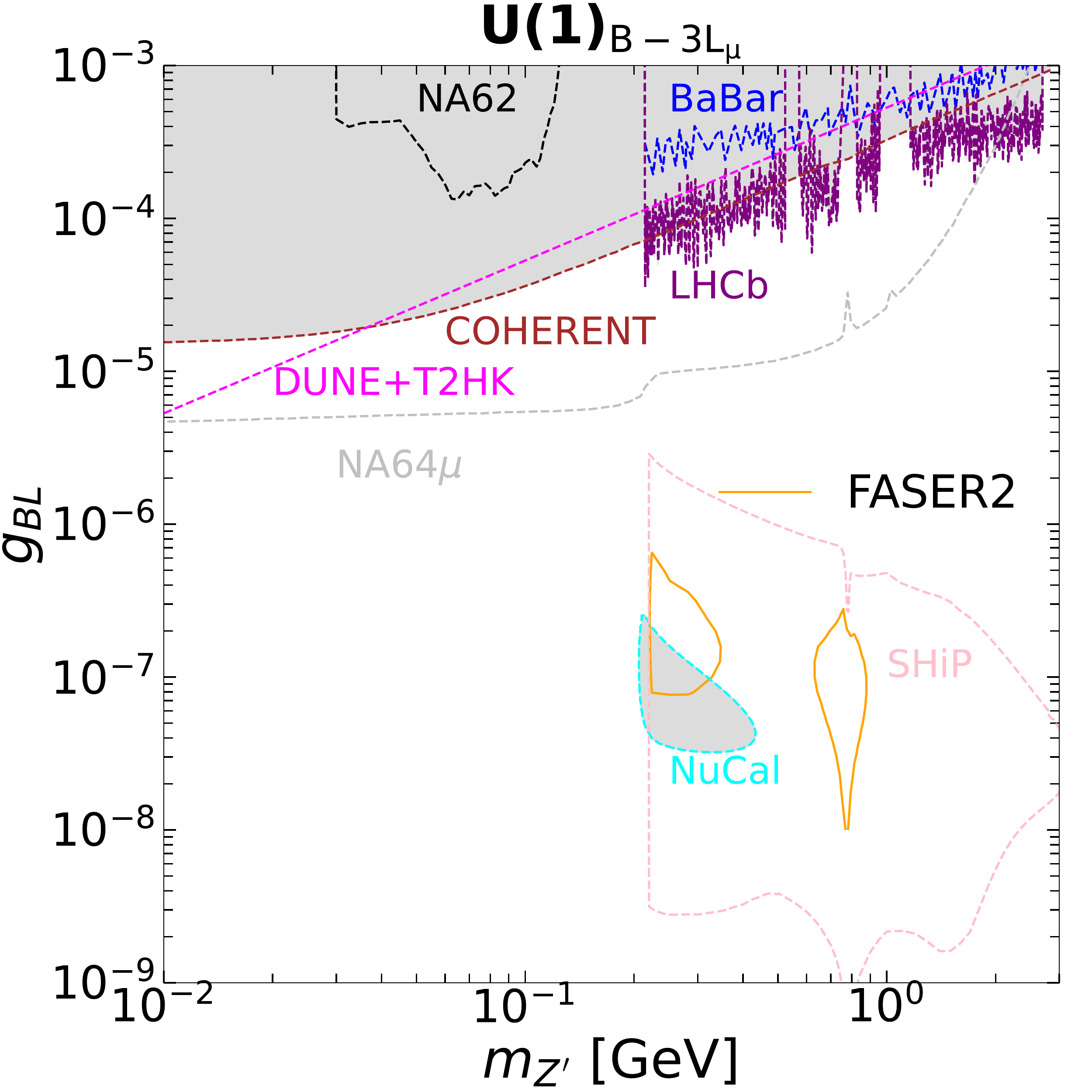}\\
\includegraphics[scale=1,width=0.45\linewidth]{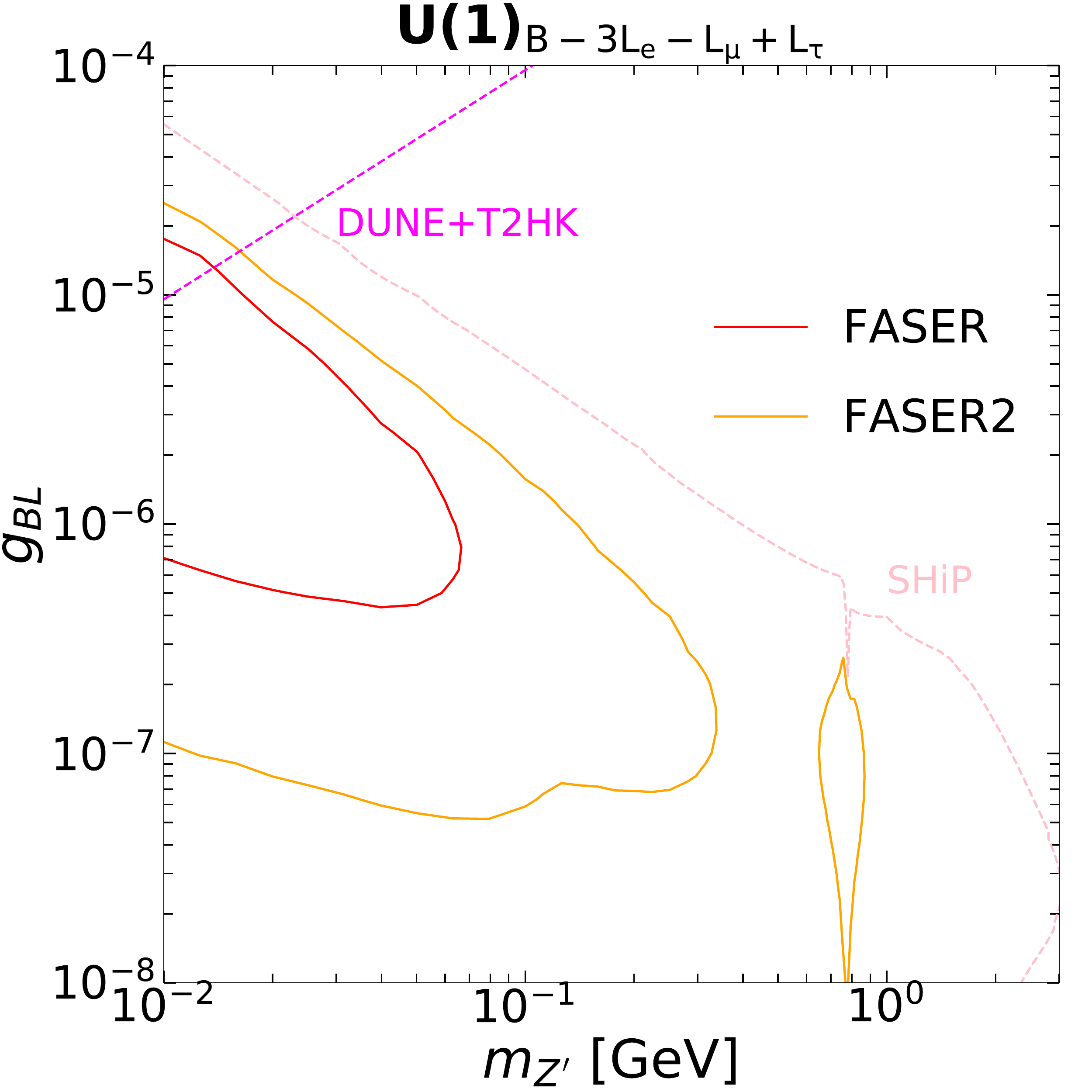}
\includegraphics[scale=1,width=0.45\linewidth]{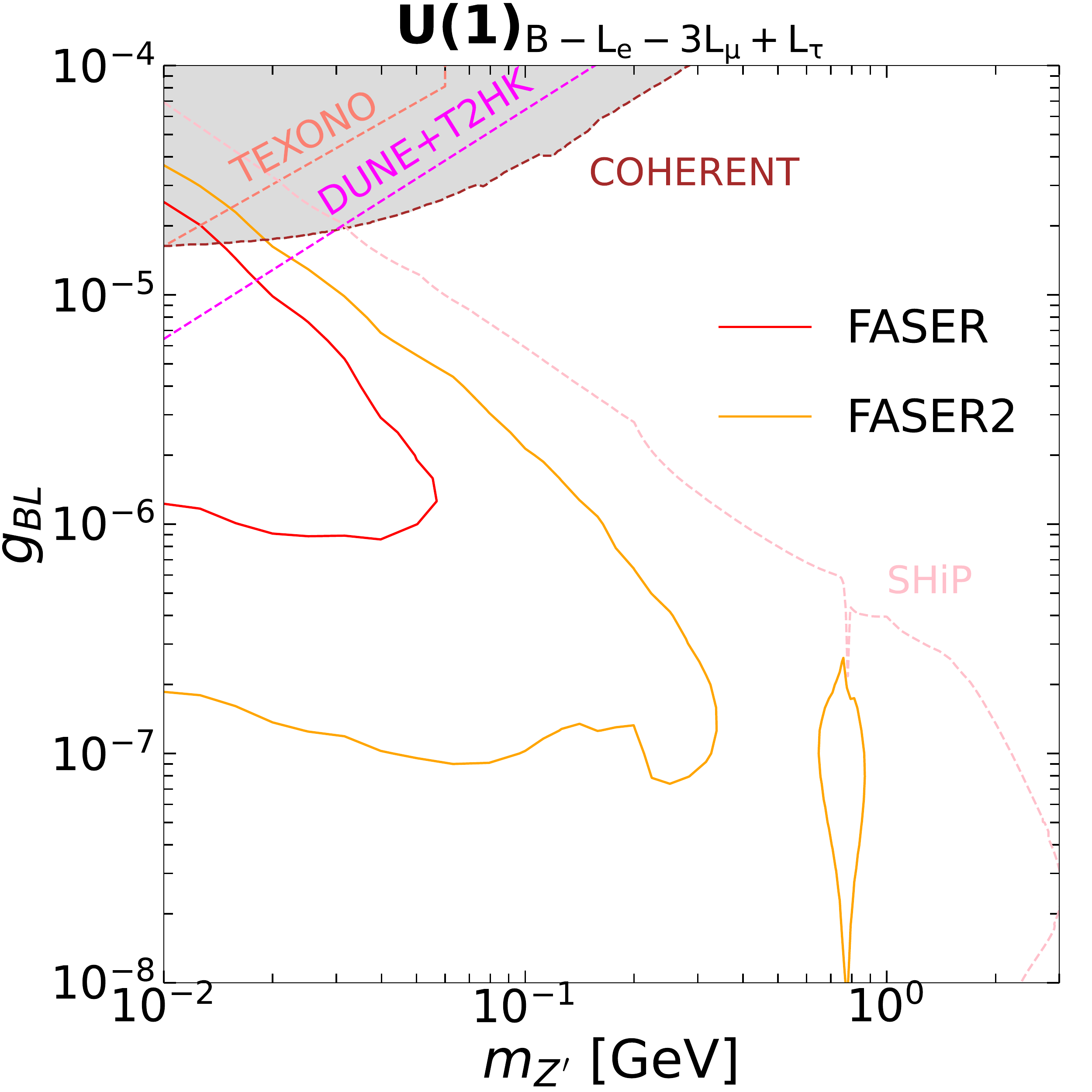}
\end{center}
\caption{The sensitivity of FASER (solid red) and FASER2 (solid orange) to the $Z'$ boson in $U(1)_{B-L}$, $U(1)_{B-3L_\mu}$, $U(1)_{B-3L_e-L_\mu+L_\tau}$ and $U(1)_{B-L_e-3L_\mu+L_\tau}$ models.
The other existing or prospective bounds for the minimal $U(1)_{B-L}$ model and $U(1)_{B-3L_\mu}$ are mainly taken from Ref.~\cite{Bauer:2018onh,Dev:2021qjj} and \cite{Bauer:2020itv}, respectively.
For $U(1)_{B-L}$, the constraints are from existing beam-dump searches~\cite{Bauer:2018onh} (purple), recent NA64e~\cite{NA64:2022yly} (silver), BaBar~\cite{BaBar:2014zli} (blue), 
CHARM-II~\cite{CHARM-II:1993phx,CHARM-II:1994dzw} (green),
TEXONO~\cite{TEXONO:2009knm} (salmon), Supernova 1987A (SN1987A)~\cite{Knapen:2017xzo} (cyan) and the DUNE Near Detector~\cite{Dev:2021qjj} (black). For $U(1)_{B-3L_\mu}$, the constraints come from NA62~\cite{NA62:2019meo} (black), BaBar~\cite{BaBar:2016sci} (blue), LHCb~\cite{LHCb:2019vmc} (purple), NA64$\mu$~\cite{Bauer:2020itv} (silver) and NuCal~\cite{Blumlein:2013cua} (cyan).
The bounds from COHERENT (dashed brown), and from the simulation of future DUNE and T2HK~\cite{Han:2019zkz} (dashed magenta) and future SHiP~\cite{SHiP:2015vad} (dashed pink) are also shown.
The SHiP 90\%CL sensitivity contour is based on the up-to-date design of SHiP as implemented in SensCalc~\cite{Ovchynnikov:2023cry}. A linear extrapolation has been used for the SHiP sensitivity contours below 21 MeV.
}
\label{fig:Zprime}
\end{figure}

For illustration we first consider the four benchmark $U(1)_{B-L}$, $U(1)_{B-3L_\mu}$, $U(1)_{B-3L_e-L_\mu+L_\tau}$ and $U(1)_{B-L_e-3L_\mu+L_\tau}$ models~\cite{Araki:2012ip,Liao:2013rca,Bhatia:2017tgo,Han:2019zkz,Bauer:2020itv,Bhatia:2021eco}.
The first one is the minimal $U(1)_{B-L}$ model with universal $U(1)'$ charge $1/3$ for SM quarks and $-1$ for leptons~\cite{Carlson:1986cu}.
The other three models with non-universal lepton charges are motivated by the introduction of vanishing textures in the neutrino mass matrix~\cite{Araki:2012ip,Liao:2013rca,Bhatia:2017tgo,Bhatia:2021eco} and have implications for non-standard neutrino interactions~\cite{Han:2019zkz,Bauer:2020itv}. The gauge boson $Z'$ in the second model only couples to the muon with $-3$ charge, and the third and fourth models have different electron and muon charges. In Fig.~\ref{fig:Zprime}, we show the sensitivity of FASER and FASER2 to the $Z'$ in these four models.
We reproduce the projected future sensitivities of FASER/FASER2 to the gauge boson $Z'$ in the minimal $U(1)_{B-L}$ model~\cite{FASER:2018eoc}.
FASER can probe the gauge coupling $g_{BL}$ down to $10^{-6}$ and $m_{Z'}$ below 0.1 GeV. FASER2 extends the sensitivity to $m_{Z'}\sim 1$ GeV and $g_{BL}\sim 10^{-7}$ for the majority of the $m_{Z'}$ region. Since it is a detector at a high-energy collider, it turns out that FASER/FASER2 is able to probe larger $g_{BL}$ than the DUNE near detector~\cite{Dev:2021qjj}\footnote{The DUNE near detector may also probe parts of the $Z'$ parameter space using $\nu-e$ scattering~\cite{Chakraborty:2021apc}.} and heavier $m_{Z'}$ compared to the beam-dump experiments~\cite{Bauer:2018onh}. Other experiments, such as NA64e~\cite{NA64:2022yly}, CHARM-II~\cite{CHARM-II:1993phx,CHARM-II:1994dzw}, BaBar~\cite{BaBar:2014zli} and TEXONO~\cite{TEXONO:2009knm}, can constrain the region of larger $g_{BL}\gtrsim 10^{-5}$. Ultimately, the proposed SHiP~\cite{SHiP:2015vad} and HIKE-dump~\cite{HIKE:2022qra} experiments will be able to further extend the reach to smaller gauge couplings, exceeding the sensitivity of FASER2.
The sensitivity of HIKE-dump (not shown in the figures) is slightly weaker than SHiP, but exceeds the sensitivity of FASER2. The less stringent constraints can be found in Refs.~\cite{Bauer:2018onh,Dev:2021qjj}.

For the $U(1)_{B-3L_\mu}$ model without electron charge, the gauge boson $Z'$ can also couple to the electron through the vacuum polarization diagram of gauge bosons~\cite{Coy:2021wfs}. However, due to the loop suppression, the coupling is proportional to $e^2/(4\pi)^2$ and has no significant impact on the search for light $Z'$ bosons. As a result, FASER/FASER2 has no sensitivity for $m_{Z'}<2m_\mu$ in this model. Other beam-dump experiments and fixed target experiments
are not sensitive to this region either. Only the search for missing energy in pion decay at NA62 is able to constrain $30\lesssim m_{Z'}\lesssim 130$ MeV. Above the dimuon threshold, FASER2 can reach isolated regions for $10^{-8}\lesssim g_{BL}\lesssim 10^{-6}$ in which there is overlap with the reach of NuCal~\cite{Merkel:2014avp} and the projection of 
SHiP~\cite{SHiP:2015vad}. Collider experiments such as BaBar~\cite{BaBar:2016sci} and LHCb~\cite{LHCb:2019vmc} exclude the region of $g_{BL}\gtrsim 10^{-4}$. The missing energy search at future NA64$\mu$~\cite{Banerjee:2019pds} provides a stronger bound for $g_{BL}\gtrsim 5\times 10^{-6}$.
Similarly to $U(1)_{B-L}$, the proposed SHiP experiment~\cite{SHiP:2015vad} will ultimately improve over the sensitivity of FASER2.
For other constraints that are less stringent or beyond the masses of interest here we refer the reader to Ref.~\cite{Bauer:2020itv}.

For the two non-universal models with electron $U(1)'$ charge, there is a blank region above $m_{Z'}\simeq 0.3$ GeV. As seen from Fig.~1 in Ref.~\cite{Ilten:2018crw}, for $m_{Z'}$ above 0.3 GeV the $\pi^+\pi^-$ channel is open and makes the $Z'$ lifetime shorter. Thus, fewer events are detected although the $\mu^+\mu^-$ decay is present. The lower panels of Fig.~\ref{fig:Zprime} show that the smaller decay rate of $\pi^+\pi^-$ also makes it difficult to detect the hadrons compared to the $U(1)_{B-L}$ model. Moreover, for the $U(1)_{B-3L_e-L_\mu+L_\tau}$ model with larger $U(1)'$ charge for electron, FASER/FASER2 extends the sensitivity to lower $g_{BL}$, which will be further extended by SHiP~\cite{SHiP:2015vad}.
The neutrino experiment TEXONO measured the elastic $\bar{\nu}_e + e^-$ scattering and its bound applies to the cases with charge $Q'_e=-1$, such as the $U(1)_{B-L_e-3L_\mu+L_\tau}$ model. As relying on interference effects of the $Z'$ with the SM, the bounds from other neutrino experiments are not straightforward to be rescaled for other models and require further work. Similarly, the beam-dump experiments measuring a very displaced vertex are also difficult to apply for these models. We leave their study for future work.

\begin{figure}[htb!]
\begin{center}
\includegraphics[scale=1,width=0.4\linewidth]{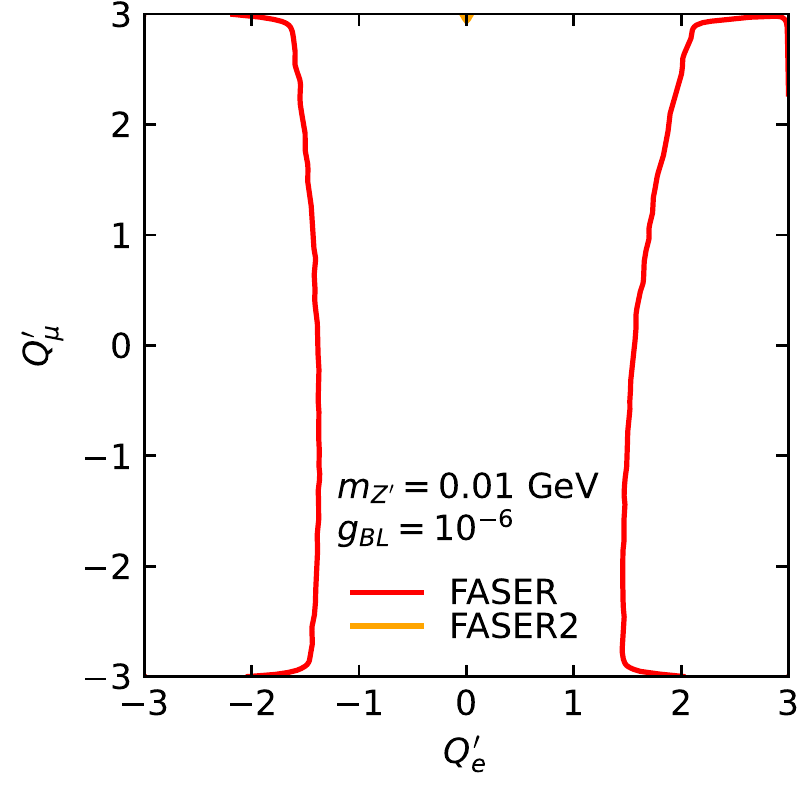}
\includegraphics[scale=1,width=0.4\linewidth]{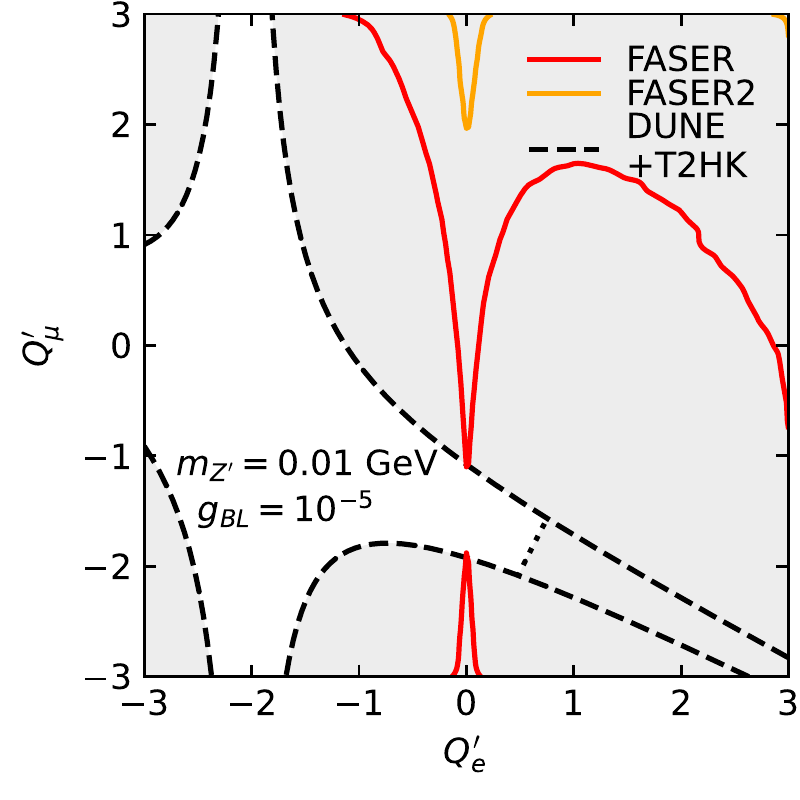}\\
\includegraphics[scale=1,width=0.4\linewidth]{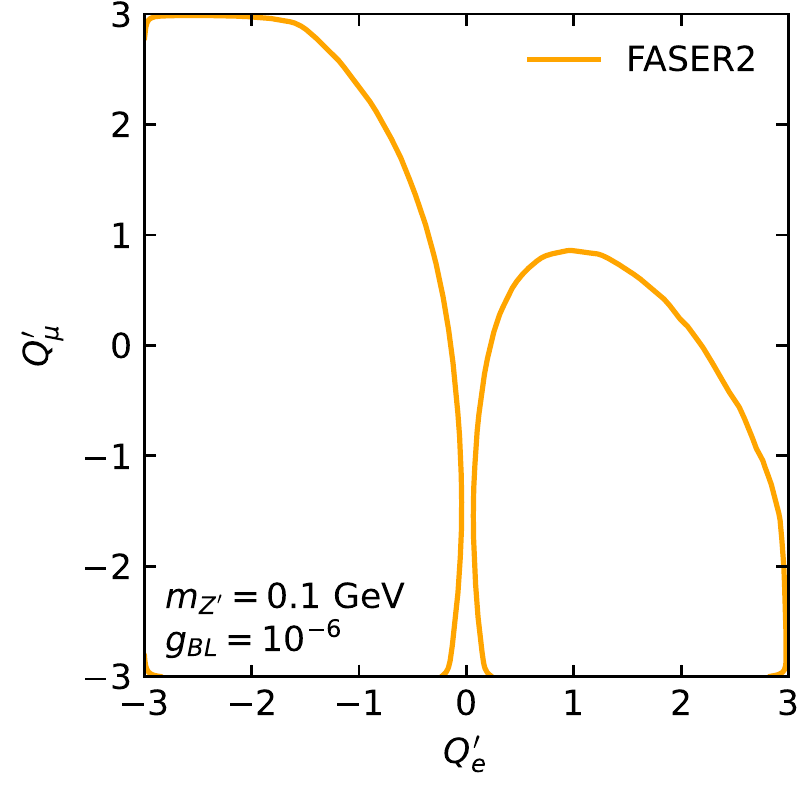}
\includegraphics[scale=1,width=0.4\linewidth]{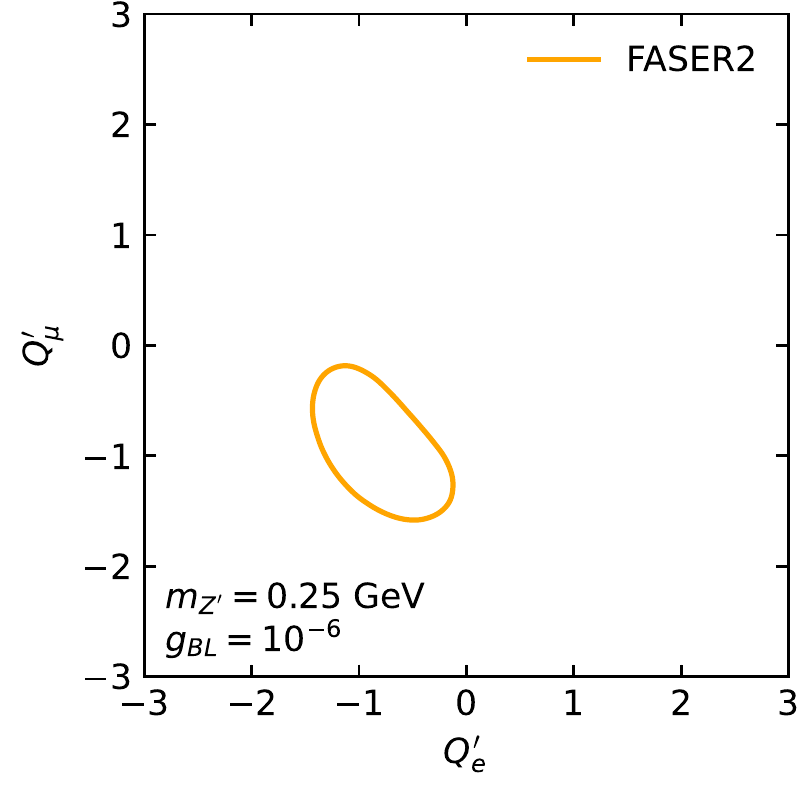}\\
\includegraphics[scale=1,width=0.4\linewidth]{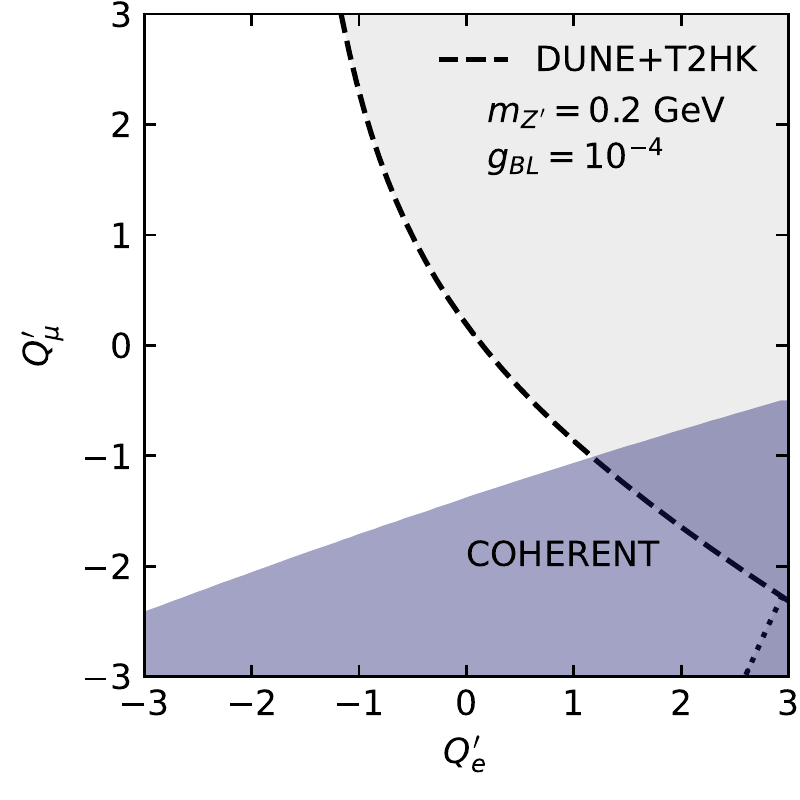}
\end{center}
\caption{FASER/FASER2's sensitivity and existing constraints from COHERENT (blue-shaded region) as function of $Q'_e$ and $Q'_\mu$ for different benchmark values of $m_{Z'}$ and $g_{BL}$. The grey regions with dashed (dotted) lines as boundaries show the parameter space which will be probed by DUNE+T2HK via $\epsilon_{\tau\tau}-\epsilon_{\mu\mu}$ ($\epsilon_{ee}-\epsilon_{\mu\mu}$), as quoted in Table~\ref{tab:osc}.}
\label{fig:QeQmu}
\end{figure}

\begin{figure}[tbp!]
\begin{center}
\includegraphics[scale=1,width=0.45\linewidth]{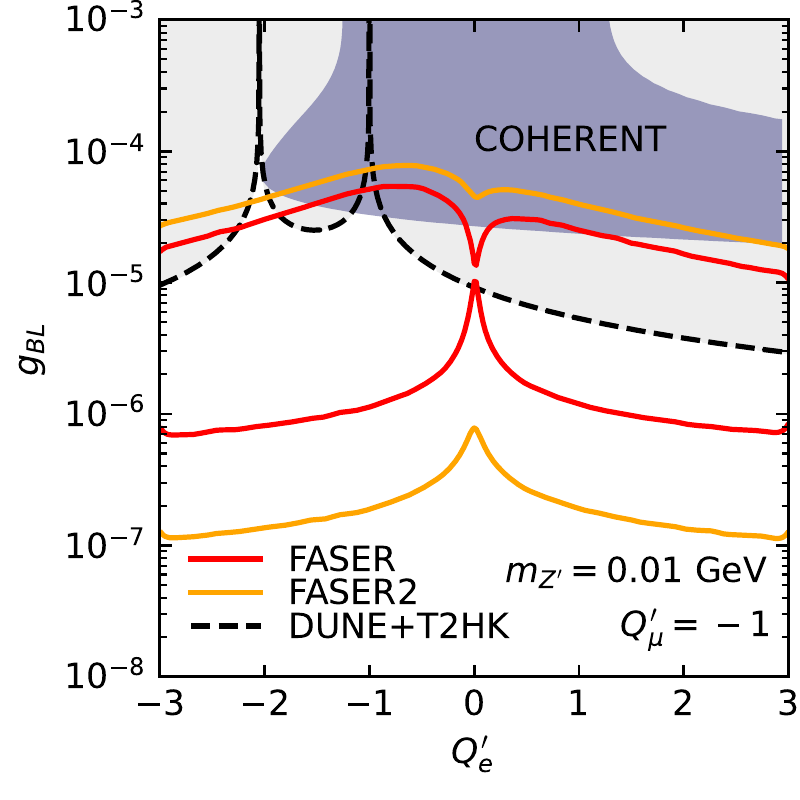}
\includegraphics[scale=1,width=0.45\linewidth]{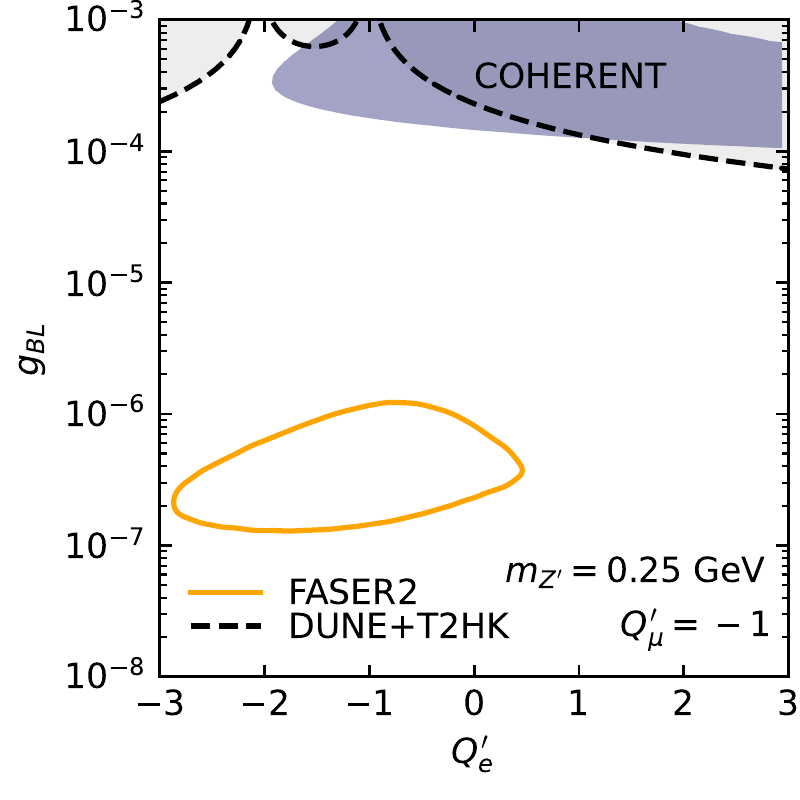}
\end{center}
\caption{FASER/FASER2's sensitivity and constraints as function of $g_{BL}$ and $Q_e^\prime$ for different benchmark values of $m_{Z'}$ and $Q_\mu^\prime$.
The grey regions with dashed lines as boundaries show the future sensitivity of DUNE+T2HK to $g_{BL}$ as a function of $Q_e'$ from the constraint on  $\epsilon_{\tau\tau}-\epsilon_{\mu\mu}$. The constraint from $\epsilon_{ee}-\epsilon_{\mu\mu}$ is weaker. The blue region shows the current constraint from COHERENT.
}
\label{fig:Qeg}
\end{figure}

In the following we focus on FASER, COHERENT and neutrino long-baseline oscillation experiments, which are either ongoing or already funded.
Next we vary two free leptonic charges $Q'_e$ and $Q'_\mu$ and show the sensitivity in the plane of these two charges in Fig.~\ref{fig:QeQmu}. FASER (FASER2) is sensitive to the regions inside the red (orange) contours. In the upper two panels we fix $m_{Z'}=0.01$ GeV and $g_{BL}=10^{-5}$ or $10^{-6}$. One can see that, as expected, FASER is sensitive to a broader region of $U(1)'$ charge for larger $g_{BL}$. $|Q'_e|>0.1~(1.5)$ can be reached when $g_{BL}=10^{-5}~(10^{-6})$. For positive $Q'_e$, FASER can probe the region of $Q'_\mu<1.5$ and the detector is sensitive to larger $Q'_\mu$ for negative $Q'_e$ because the tau lepton is assigned smaller $|Q'_\tau|$. FASER2 can reach the whole region of $|Q'_e|>0.1$ and has no dependence on $Q'_\mu$. In the lower two panels we fix $g_{BL}=10^{-6}$ and $m_{Z'}=0.1$ GeV or 0.25 GeV. FASER has no sensitivity for these larger $m_{Z'}$ values. FASER2 can cover a larger region for smaller $m_{Z'}$ and only leaves an island of sensitivity in the plane of these two charges for $m_{Z'}=0.25$ GeV.

Finally, in Fig.~\ref{fig:Qeg} we show the constraints as function of $g_{BL}$ and $Q'_e$ for different benchmark values of $m_{Z'}$ and $Q'_\mu$.
For small $Z'$ mass of 0.01 GeV, FASER (FASER2) can place a constraint on $g_{BL}$ down to $10^{-6}~(10^{-7})$ as long as $|Q'_e|>0.1$. For a large value of $m_{Z'}=0.25$ GeV, however, only FASER2 can cover a narrow region of $10^{-7}<g_{BL}<10^{-6}$.

\subsection{The search reach of COHERENT and neutrino oscillation for \texorpdfstring{$Z'$}{Z'}}
While FASER and FASER2 are far-forward detectors which are sensitive to long-lived particles and thus small gauge couplings $g_{BL}$, COHERENT and neutrino oscillation experiments probe larger gauge couplings $g_{BL}$. Thus COHERENT and neutrino oscillation experiments provide complementary sensitivity to FASER and other far-forward detectors.

\subsubsection{COHERENT}

In order to account for the latest data release on the measurement of coherent elastic neutrino-nucleus scattering (CE$\nu$NS) in CsI by COHERENT~\cite{COHERENT:2021xmm}, we employ the Poissonian least-squares function
\begin{equation}\label{eq:LSF}
		\chi^2 = 2\sum_{i=2}^{9}\sum_{j=1}^{11}\left[N^{\text{th}}_{ij} - N^{\text{exp}}_{ij} + N^{\text{exp}}_{ij}\ln\frac{N^{\text{exp}}_{ij}}{N^{\text{th}}_{ij}}\right] + \left(\frac{\alpha_0}{\sigma_0}\right)^2 + \left(\frac{\alpha_3}{\sigma_3}\right)^2
\end{equation}
with
\begin{equation}
		N^{\text{th}}_{ij} = (1 + \alpha_0)N^{\text{CE$\nu$NS}}_{ij}(\alpha_4,\alpha_6,\alpha_7) + (1 + \alpha_3)N^{\text{SSB}}_{ij}
\end{equation}
as well as $\sigma_0 = 0.1145$ and $\sigma_3 = 0.021$. For more information on how to obtain the bin-wise signal prediction $N^{\text{CE$\nu$NS}}_{ij}$ from integrating over the relevant differential cross sections and flux distributions, we refer the reader to Ref.~\cite{DeRomeri:2022twg}, including the references therein, that our analysis is closely aligned with. Several systematic uncertainties are captured via the nuisance parameters $\alpha_i$, the numerical values of which are unknown a priori and thus must be fitted from the data. The above indicates that we only include the beam-uncorrelated steady-state background (SSB) which constitutes over 98\% of the background as is reported in~\cite{COHERENT:2021xmm}.
We also follow their distribution into energy bins, expressed in photoelectrons, with the endpoints
\begin{equation}
    E \sim (8,12,16,20,24,32,40,48,60)~\text{PE}
\end{equation}
and recoil time bins with the endpoints
\begin{equation}
    t_{\text{rec}}\sim (0,0.125,0.25,0.375,0.5,0.625,0.75,0.875,1,2,4,6)~\mu\text{s} .
\end{equation}
We do not include the lowest-energy bin $E\sim(0,8)~\text{PE}$ in the analysis, because the used detector efficiency function is negative in that region. The contribution from the $Z'$ boson to CE$\nu$NS is induced via a modification of the effective weak charge of the nucleus (see for instance Ref.~\cite{AtzoriCorona:2022moj}),
\begin{align}\label{eq:weak_charge_nucleus}
	Q_{\nu,k} & = Q^{\text{SM}}_{i,k} + Q^{Z'}_{i,k}   = g_{p,i}Z_{0,k} + g_{n,i}N_{0,k} + \frac{
    g_{BL}^2
    Q'_i\left(Z_{0,k} + N_{0,k} \right)}{\sqrt{2}G_F\left(|\mathbf{q}|^2 + m_{Z'}^2\right)}
\end{align}
with the three neutrino species $\nu_i \in \nu_e, \nu_\mu, \overline{\nu}_\mu$ and the two nuclei $k = \text{Cs}, \text{I}$, and $\mathbf{q}$ denotes the 3-momentum of the virtual $Z'$ boson.
We have explicitly checked that the contributions from neutrino-electron scattering are entirely negligible in the considered mass range $m_{Z'}\gtrsim 0.01$ GeV.
The contour lines visible in Figs.~\ref{fig:Zprime}, \ref{fig:QeQmu} and \ref{fig:Qeg} arise from bounding the profile $\Delta\chi^2 \equiv \chi^2 - \chi^2_{\text{min}}$ from above, where for each (fixed) set of model parameters ($Q'_e, Q'_\mu, g_{BL},m_{Z'}$) we choose the nuisance parameters $\alpha_i$ such that $\chi^2$ is minimised, and $\chi^2_{\text{min}}$ is the (global) minimum value taken when both the model parameters and the nuisance parameters are optimised simultaneously. The confidence level is set to 90\%. As can be seen from Eq.~(\ref{eq:weak_charge_nucleus}), there is a degeneracy in the model parameters since we are not sensitive to the gauge coupling and charges individually, but only to the product $g_{BL}^2Q_i'$.

For the $Z'$ mass and gauge coupling fixed, we find that the following two non-SM-like cases are generally in good agreement with the COHERENT data: (i) $Q_\mu' > 0$ for $|Q_\mu'|$ large, and (ii) $Q_e' < 0$ for $|Q_e'|$ large and, if $Q_\mu'<0$, $|Q_\mu'|$ not too large, as can be seen in the bottom plot of Fig.~\ref{fig:QeQmu}. We interpret this as an indication that besides SM-like cases, there is also a preference for negative corrections to $Q_{i,k}^{\text{SM}}$.
In the $U(1)_{B-L}$ model, for a range of $Z'$ masses this can be realised if the $g_{BL}$ gauge coupling becomes large enough. Indeed, the lower tip of a non-excluded region for $g_{BL}
\gtrsim 10^{-3}$ is visible the top-left plot of Fig.~\ref{fig:Zprime}. This is in agreement with Fig.~\ref{fig:Qeg} which demonstrates that for the benchmark case $Q_\mu' = -1$ and a moderately large $|Q_e'|$, a specific region of larger $g_{BL}$
can exist that is less constrained than $g_{BL}$
smaller. The $U(1)_{B-3L_e-L_\mu+L_\tau}$ model represents a more extreme case to which COHERENT does not provide any sensitivity and thus there is no constraint at all. Lastly, we comment that the slope of the lower bound of the excluded region for the $U(1)_{B-L}$, $U(1)_{B-3L_\mu}$ and $U(1)_{B-L_e-3L_\mu+L_\tau}$ models approaches a constant value for larger $Z'$ masses, consistent with the expectation that the contribution to CE$\nu$NS will approximately scale like $g_{BL}^2/m_{Z'}^2$ in that regime.

\subsubsection{Neutrino oscillation experiments}

The Hamiltonian for neutrino propagation in matter
in the presence of a $U(1)'$ model becomes
\begin{eqnarray}
H={1\over 2E}U \left(
  \begin{array}{ccc}
   0  & 0 & 0 \\
   0  & \delta m_{21}^2 & 0 \\
   0  & 0 & \delta m_{31}^2 \\
  \end{array}
\right) U^\dagger +
V_{CC} \left(
  \begin{array}{ccc}
   1+\epsilon_{ee}  & \epsilon_{e\mu} & \epsilon_{e\tau} \\
   \epsilon_{e\mu}^\ast & \epsilon_{\mu\mu} & \epsilon_{\mu\tau} \\
   \epsilon_{e\tau}^\ast & \epsilon_{\mu\tau}^\ast & \epsilon_{\tau\tau}
   \\
  \end{array}
\right)\;,
\end{eqnarray}
where $U$ is the Pontecorvo-Maki-Nakagawa-Sakata (PMNS) matrix, $V_{CC}=\sqrt{2}G_F N_e$ is the standard matter potential, and $\epsilon_{\alpha\beta}=\sum_f\epsilon^f_{\alpha\beta} N_f/N_e$ with $N_{f}$ being the number density of fermions $f=u,d$ and $e$. The quark number densities satisfy
\begin{eqnarray}
N_u=2N_p+N_n\;,~~N_d=N_p+2N_n\;,
\end{eqnarray}
where $N_p$ and $N_n$ are proton and neutron number densities, respectively, and the matter neutrality implies $N_p=N_e$.
The diagonal parameters of non-standard neutrino interactions (NSI) are given by~\cite{Han:2019zkz}
\begin{eqnarray}
\epsilon_{\alpha\alpha}^f={g_{BL}^2 Q'_\alpha Q'_f\over \sqrt{2}G_F m_{Z'}^2}\;,
\end{eqnarray}
where $\alpha$ denotes charged lepton flavor.
Note that there are no off-diagonal terms, since the theory is defined in the basis where the charged lepton masses are diagonal.
As a result, in our general $U(1)'$ models, the diagonal components of the matter part in the Hamiltonian are
\begin{eqnarray}
\epsilon_{\alpha\alpha}={g_{BL}^2 \over \sqrt{2}G_F m_{Z'}^2}\Big[Q'_\alpha Q'_e+Q'_\alpha Q'_q\Big(3+3{N_n\over N_e}\Big)\Big]\;,
\end{eqnarray}
where $N_n/N_e\approx 1.051$ in Earth matter.

In Table~\ref{tab:osc}, we show the $2\sigma$ allowed ranges for the diagonal NSI parameters from the simulation of future long-baseline experiments DUNE and T2HK~\cite{Liao:2016orc, Han:2019zkz}. The neutrino oscillation data constrain differences between two diagonal $\epsilon$ parameters. Thus, there is no bound on the $U(1)_{B-L}$ model from neutrino oscillation, but only on those with non-universal lepton charges as shown in Fig.~\ref{fig:Zprime}. Note that due to the presence of the generalized mass ordering degeneracy~\cite{Coloma:2016gei}, the allowed region near $\epsilon_{ee}-\epsilon_{\mu\mu}=-2$ cannot be ruled out by neutrino oscillation experiments alone.
In the literature studying the current neutrino oscillation data~\cite{Esteban:2018ppq,Coloma:2023ixt,Amaral:2023tbs}, there exist bounds on only electron NSI couplings $\epsilon^e_{\alpha\alpha}$, only quark couplings $\epsilon^q_{\alpha\alpha}$, or components $\epsilon_{\alpha\alpha}$ from a global analysis including CE$\nu$NS data within an effective theory and thus only applicable for $m_{Z'}>50$ MeV. We thus do not consider the constraints from the current neutrino oscillation data here.


As seen in Fig.~\ref{fig:Zprime}, DUNE+T2HK will be able to probe lower $g_{BL}$ than COHERENT for $m_{Z'}<30$ MeV in $U(1)_{B-3L_\mu}$ and $U(1)_{B-L_e-3L_\mu+L_\tau}$ models. For the $U(1)_{B-3L_e-L_\mu+L_\tau}$ model in which COHERENT loses sensitivity, they can also probe the regime of large $g_{BL}$ and small $m_{Z'}$. In Figs.~\ref{fig:QeQmu} and \ref{fig:Qeg}, the grey regions show the future sensitivity of DUNE+T2HK. They provide complementary projected bounds compared to FASER/FASER2 and COHERENT. It turns out that the $\epsilon_{\tau\tau}-\epsilon_{\mu\mu}$ quantity can place stronger bounds. The two ``peaks'' losing sensitivity from left to right in Fig.~\ref{fig:Qeg} correspond to $\epsilon_{\alpha\alpha}=0$ and $\epsilon_{ee}-\epsilon_{\mu\mu}=\epsilon_{\tau\tau}-\epsilon_{\mu\mu}=0$, respectively.

\begin{table}
\begin{center}
\begin{tabular}{c|c}
\hline
   & DUNE+T2HK \\\hline
$\epsilon_{ee}-\epsilon_{\mu\mu}$& $[-2.50,-1.66]\oplus [-0.443,+0.394]$ \\
\hline
$\epsilon_{\tau\tau}-\epsilon_{\mu\mu}$ & $[-0.105,+0.105]$ \\
\hline
\end{tabular}
\end{center}
\caption{The $2\sigma$ allowed ranges for the diagonal NSI parameters from the simulation of future DUNE and T2HK~\cite{Liao:2016orc, Han:2019zkz}.}
\label{tab:osc}
\end{table}

\section{Sensitivity of FASER to scalar \texorpdfstring{$\phi$}{f} and increased
\texorpdfstring{$Z'$}{Z'} production}
\label{sec:FASERphi}

\subsection{Sensitivity of FASER to scalar \texorpdfstring{$\phi$}{f}}
\label{sec:FASERphionly}

The interactions of the scalar $\phi$ can be obtained via the mixing with the SM Higgs boson. It can also become a light and weakly interacting particle governed by the small mixing angle $\theta$~\cite{Winkler:2018qyg}. Below we will explore the sensitivity of FASER/FASER2 to light scalars $\phi$.

The scalar $\phi$ has flavour-changing interactions via its mixing with the SM Higgs boson. For down-type quarks they are described by the effective Lagrangian~\cite{Willey:1982mc,Leutwyler:1989xj}\footnote{See \cite{Kachanovich:2020yhi} for a calculation of the effective vertex of the second Higgs in $R_\xi$ gauge.}
\begin{equation}\label{eq:FCNCcoupling}
    \mathcal{L}_{\rm FCNC} =  \frac{-\phi}{v_0} \sin\theta \sum_{\alpha,\beta} C_{\alpha\beta}  \left(m_{d_\beta} \, \bar d_\alpha P_R d_\beta\, \right) + \mathrm{h.c.} \qquad\mathrm{with} \quad C_{\alpha\beta}
    \approx  \frac{3 m_t^2 V_{t\alpha}^\ast V_{t\beta}}{16\pi^2 v_0^2} \;.
\end{equation}
Normally, the light scalar $\phi$ is mainly produced in the flavor changing 2-body rare decay of $B$ mesons $b\to X_s\phi$. This process includes all b-flavored hadrons and all strange-flavored decay products. The corresponding branching fraction is given by~\cite{Feng:2017vli}
\begin{eqnarray}
{\rm BR}(b\to X_s \phi)=5.7\times \Big(1-{m_\phi^2\over m_b^2}\Big)^2 \times \sin^2\theta\;.
\end{eqnarray}
Moreover, the Kaons are expected to be produced more copiously than $B$ mesons at the LHC.
The light scalar $\phi$ can also be produced from Kaon decays $K^\pm\to \pi^\pm \phi$ and $K_{L(S)}\to \pi^0 \phi$~\cite{Leutwyler:1989xj,Feng:2017vli,Dev:2019hho,Batell:2020vqn,Dev:2021qjj,Li:2022zgr}.
The amplitude of the Kaon decay into the light scalar receives contributions from the effective couplings of the scalar to the gluon field strength tensor which are proportional to the coupling constants $\gamma_1,\gamma_2$ and the effective FCNC coupling of the scalar in Eq.~\eqref{eq:FCNCcoupling}~\cite{Leutwyler:1989xj}. It takes the form
\begin{eqnarray}
\mathcal{M}={-\sin\theta \over v_0}\Big[\gamma_1{7\over 18}(m_K^2-m_\phi^2+m_\pi^2)-\gamma_2{7\over 9}m_K^2+ \frac12 C_{ds} m_s{m_K^2-m_\pi^2\over m_s-m_d}f_0^{K\pi} \Big]\;,
\end{eqnarray}
where $\gamma_1\approx 3.1\times 10^{-7}$, $\gamma_2\approx 0$~\cite{Leutwyler:1989xj} and the form factor $f_0^{K\pi}= 0.96$~\cite{Boiarska:2019jym}. After only taking the dominant third term in the amplitude, the branching fractions of Kaon decays become
\begin{align}
{\rm BR}(K^\pm\to \pi^\pm\phi)&= \frac{\lambda^{1/2}(m_K^2,m_\pi^2,m_\phi^2)}{16\pi m_K^3 \Gamma_{K^\pm}} |\mathcal{M}|^2\approx 1.7\times 10^{-3}\lambda^{1/2}\left(1,\frac{m_\pi^2}{m_K^2},\frac{m_\phi^2}{m_K^2}\right)\sin^2\theta\;,\\
{\rm BR}(K_L\to \pi^0\phi)&= \frac{\lambda^{1/2}(m_K^2,m_\pi^2,m_\phi^2)}{16\pi m_K^3 \Gamma_{K_L}} (\mathrm{Re}\mathcal{M})^2\approx 6.1\times 10^{-3}\lambda^{1/2}\left(1,\frac{m_\pi^2}{m_K^2},\frac{m_\phi^2}{m_K^2}\right)\sin^2\theta\;,\\
{\rm BR}(K_S\to \pi^0\phi)&= \frac{\lambda^{1/2}(m_K^2,m_\pi^2,m_\phi^2)}{16\pi m_K^3 \Gamma_{K_S}} (\mathrm{Im}\mathcal{M})^2\approx 2.0\times 10^{-6}\lambda^{1/2}\left(1,\frac{m_\pi^2}{m_K^2},\frac{m_\phi^2}{m_K^2}\right)\sin^2\theta
\end{align}
with the K\"all\'en function $\lambda(a,b,c)=a^2+b^2+c^2-2(ab+ac+bc)$.
The Kaon decays contribute to the production of lighter $\phi$ which is more severely constrained from other experiments.
We include this production mode for completeness. The branching ratio for  Higgs decay $h\to \phi\phi$ is BR($h\to \phi\phi) \approx 0.19 \, (500 \,\mathrm{GeV} Q_\Phi' g_{BL}/m_{Z'})^2 (\sin\theta/0.1)^2$ and thus strongly constrained by searches for the $Z'$ gauge boson in most of the available parameter space, see e.g.~\cite{Kling:2020iar} for a compilation of many existing constraints.
Thus, for simplicity, we neglect the production of the light scalar $\phi$ from SM Higgs decay.

\begin{figure}[tb!]
\begin{center}
\includegraphics[scale=1,width=0.47\linewidth]{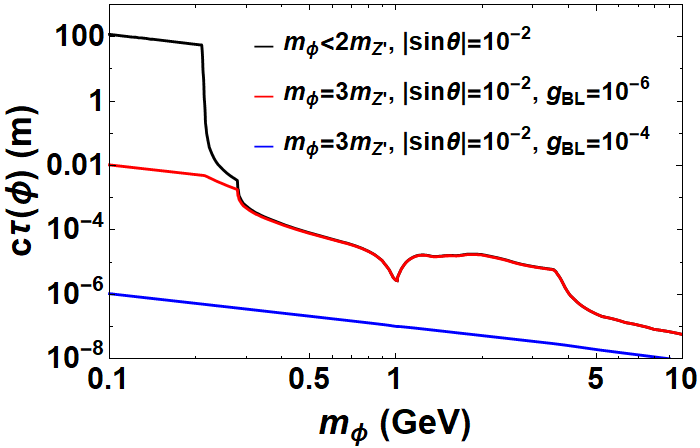}
\includegraphics[scale=1,width=0.47\linewidth]{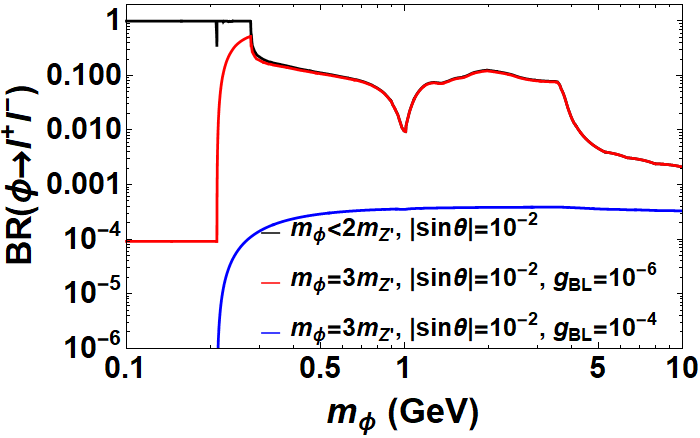}
\end{center}
\caption{The decay length (left) and decay branching ratios of $\phi$ into $e^+e^-$ and $\mu^+\mu^-$ (right), for $m_\phi<2 m_{Z'}$ with $|\sin\theta|=10^{-2}$ (black) and $m_\phi=3m_{Z'}$ with $g_{BL}=10^{-6}$ (red) or $g_{BL}=10^{-4}$ (blue).
}
\label{fig:brphi}
\end{figure}

The decays of light $\phi$ into SM matters actually include $\phi\to \ell^+\ell^-$ ($\ell=e,\mu$), $\gamma\gamma$ and $\pi\pi$ for $m_\phi<1$ GeV~\cite{Dev:2021qjj}.
The partial decay widths are then given by
\begin{eqnarray}
\Gamma(\phi\to \ell^+\ell^-)&=&{G_F m_\phi m_\ell^2 \sin^2\theta\over 4\sqrt{2}\pi}\Big(1-{4m_\ell^2\over m_\phi^2}\Big)^{3/2}\;,\\
\Gamma(\phi\to \pi^+\pi^-)&=&2\Gamma(\phi\to \pi^0\pi^0)={G_F m_\phi^3 \sin^2\theta\over 8\sqrt{2}\pi}\Big(1-{4m_\pi^2\over m_\phi^2}\Big)^{1/2}|G(m_\phi^2)|^2\;,\\
\Gamma(\phi\to \gamma\gamma)&=&{G_F \alpha^2 m_\phi^3 \sin^2\theta\over 128\sqrt{2}\pi^3}\Big|\sum_f C_f Q_f^2 A_{1/2}(\tau_f)+A_1(\tau_W)\Big|^2\;,
\end{eqnarray}
where $G(m_\phi^2)$ is a dimensionless transition amplitude~\cite{Donoghue:1990xh}, $A_{1/2}(\tau_f)$ and $A_1(\tau_W)$ with $\tau_X=m_\phi^2/4m_X^2$ are the standard
fermion and $W$ loop functions for the SM Higgs decay~\cite{Djouadi:2005gi}, as summarized in the Appendix of Ref.~\cite{Dev:2021qjj}.
When $m_\phi\gtrsim 1$ GeV, the $KK$ decay mode is allowed.
Note that hadronic scalar decays have a large theory uncertainty. See~\cite{Winkler:2018qyg,Gorbunov:2023lga} for two recent independent calculations for the hadronic decays. As hadronic scalar decays dominate the total decay width above the pion threshold, the scalar lifetime has a similar uncertainty.
In our numerical calculation below, we adopt the default inputs of the dark Higgs model in FORESEE for the scalar decay into SM matter fields.
Moreover, when $m_\phi>2m_{Z'}$, the scalar can also decay into a pair of $Z'$ boson from the covariant derivative term $(D_\mu \Phi)^\dagger (D^\mu \Phi)$. The corresponding decay width is determined by the gauge coupling $g_{BL}$ as follows~\footnote{We find that the decay width is larger by a factor 4 compared to Eq.~(2.8) in Ref.~\cite{Dev:2021qjj}.}
\begin{eqnarray}\label{eq:phiZpZp}
\Gamma(\phi\to Z'Z')={g_{BL}^2(Q'_\Phi)^2m_\phi^3\over 32\pi m_{Z'}^2}\Big(1-{4m_{Z'}^2\over m_\phi^2}\Big)^{1/2}\Big(1-{4m_{Z'}^2\over m_\phi^2}+{12m_{Z'}^4\over m_\phi^4}\Big)\;.
\end{eqnarray}
Thus, there are two additional
parameters $m_{Z'}$ and $g_{BL}$ in this case.
In Fig.~\ref{fig:brphi} we show the total decay length of $\phi$ (left) and the branching ratios (right) of $\phi$ decay into $\ell^+\ell^-$ ($\ell=e$ and $\mu$) with or without the decay $\phi\to Z'Z'$. Without $\phi\to Z'Z'$, the $e^+e^-$ channel is dominant for $m_\phi<0.2$ GeV before the $\mu^+\mu^-$ channel is kinematically allowed. Given $\phi\to Z'Z'$ is allowed with $m_\phi=3m_{Z'}$ and $g_{BL}=10^{-6}$, in the region of $m_\phi<2m_\mu$, the total decay length and BR$(\phi\to e^+ e^-)$ are both suppressed by four orders of magnitude. When $g_{BL}$ becomes large enough, $\phi\to Z'Z'$ dominates over all decay channels into SM products and the decay length decreases linearly with $m_\phi$.

\begin{figure}[tb!]
\begin{center}
\includegraphics[scale=1,width=0.45\linewidth]{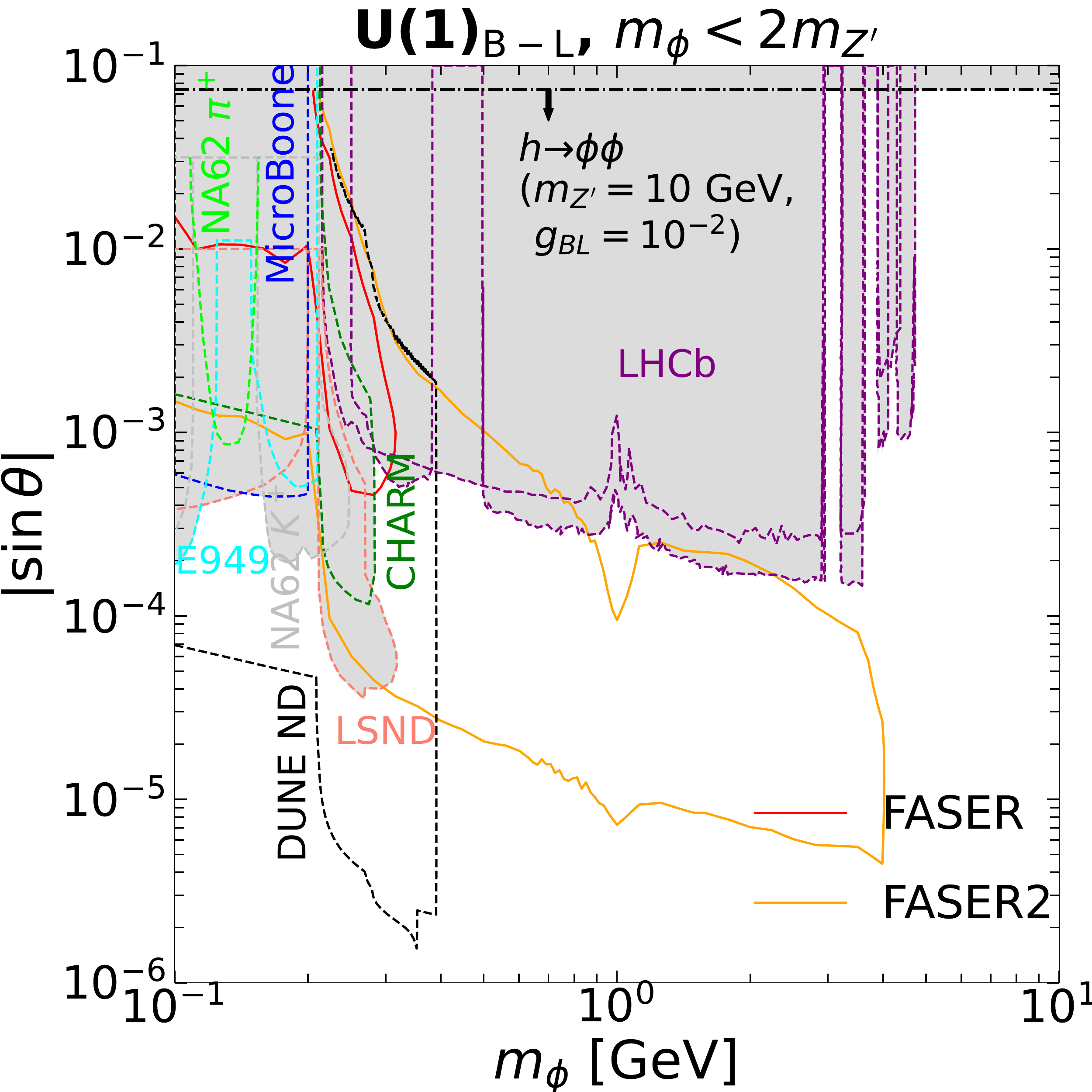}\\
\includegraphics[scale=1,width=0.45\linewidth]{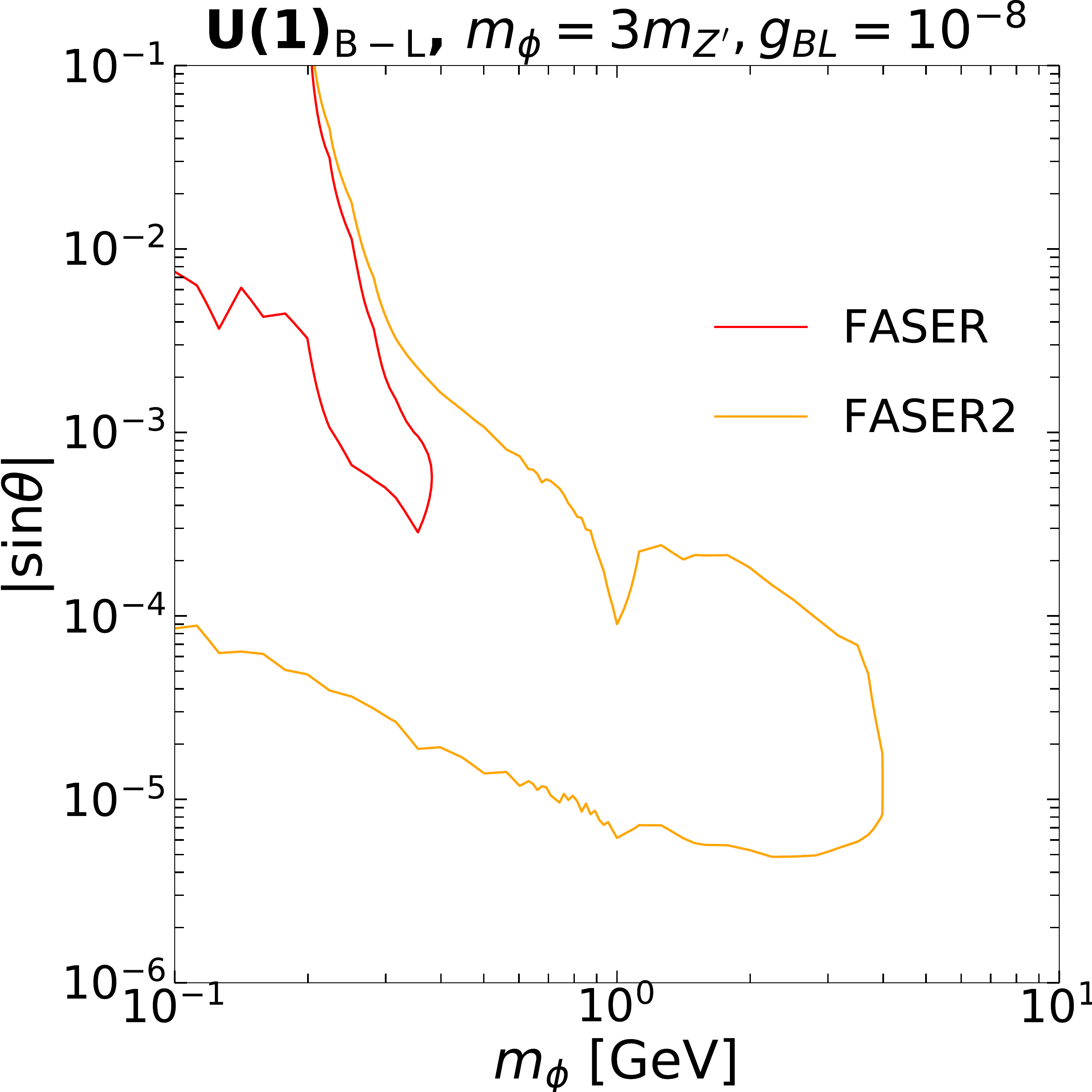}
\includegraphics[scale=1,width=0.45\linewidth]{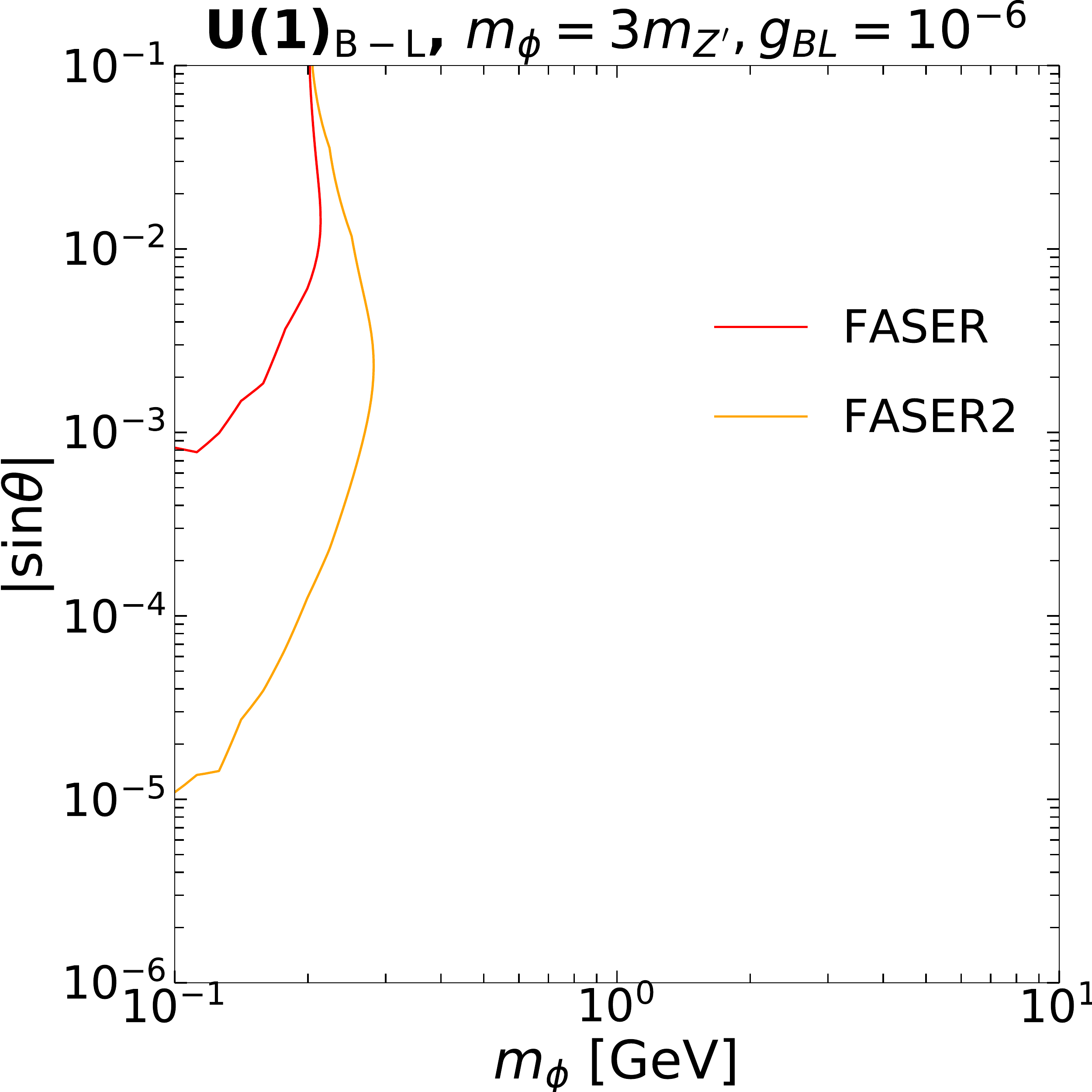}
\end{center}
\caption{The sensitivity of FASER and FASER2 to the mixing $\sin\theta$ of the scalar $\phi$ in case A ($m_\phi<2 m_{Z'}$) and case B ($m_\phi=3 m_{Z'}$, $g_{BL}=10^{-8}$ or $g_{BL}=10^{-6}$).
The other existing bounds for the $m_\phi<2m_{Z'}$ case are taken from Refs.~\cite{Kling:2021fwx} and \cite{Dev:2021qjj}, including NA62~\cite{Ruggiero:2020phq} (light green), CHARM~\cite{Winkler:2018qyg} (green), MicroBooNE~\cite{MicroBooNE:2021usw} (blue), BNL-E949~\cite{BNL-E949:2009dza} (cyan), LSND~\cite{Foroughi-Abari:2020gju} (salmon), LHCb~\cite{LHCb:2015nkv,LHCb:2016awg} (purple), DUNE near detector~\cite{Dev:2021qjj} (black), and Higgs decay to scalars (dash-dotted black) as described in Eq.~\eqref{eq:Hphiphi}.
}
\label{fig:phi}
\end{figure}

To evaluate the sensitivity of FASER/FASER2 to the scalar $\phi$, we consider two cases: $m_\phi<2 m_{Z'}$ (case A) and $m_\phi>2 m_{Z'}$ (case B).
The FORESEE package is then modified to take into account the production of $\phi$ from both $B$ and $K$ meson decay and the two-dimensional parameterization of $\sin\theta$ and $m_\phi$ for the $m_\phi>2 m_{Z'}$ case.
In Fig.~\ref{fig:phi} we show the sensitivity of FASER/FASER2 to the mixing $\sin\theta$ for the scalar $\phi$ in the plane of $\sin\theta$ vs. $m_\phi$. FASER2 yields the most sensitive constraint for $m_\phi\gtrsim 0.4$ GeV as the viable decay products such as $\mu^+\mu^-$ or $\pi\pi$ are dominant once kinematically allowed.
For the case of $m_\phi<2 m_{Z'}$, FASER and FASER2 can respectively reach $\sin\theta\sim 6\times 10^{-4}$ and $5\times 10^{-6}$ at most. For lower $m_\phi$, the constraint becomes weaker to the level of $\sin\theta\sim 10^{-2}$ and $10^{-3}$ as only the $\phi\to e^+e^-$ channel dominates.
For the case B with $m_\phi>2 m_{Z'}$, we fix $m_\phi=3m_{Z'}$ and $g_{BL}=10^{-8}$ or $g_{BL}=10^{-6}$ for illustration. As seen from the lower panels of Fig.~\ref{fig:phi}, the projected constraints change dramatically as $g_{BL}$ increases. For $m_\phi<2m_\mu$, as $g_{BL}$ increases, the decay length decreases and a lower range of $\sin\theta$ can be constrained. Given $m_\phi=0.1$ GeV and $g_{BL}=10^{-6}$, a mixing $\sin\theta$ as small as $10^{-5}$ can be reached.

Note that the invisible Higgs decay could place additional constraints when $h\to \phi\phi$
or $h\to Z' Z'$ decay is allowed. Based on our Eq.~(\ref{eq:Hphiphi}), we find that there is no excluded region by $h\to \phi\phi$ for the lower panels of Fig.~\ref{fig:phi} with $m_\phi=3m_{Z'}$ and very small $g_{BL}$. For the case of $m_\phi<2m_{Z'}$ in the upper panel of Fig.~\ref{fig:phi}, the constraint depends on free parameters $m_{Z'}$ and $g_{BL}$. We take $m_{Z'}=10$ GeV and $g_{BL}=10^{-2}~(10^{-3})$ for illustration and find the upper bound as $|\sin\theta|<0.074~(0.18)$. We add the limit of $|\sin\theta|<0.074$ in the upper panel of Fig.~\ref{fig:phi} as an example.
Due to the mixing between new scalar $\phi$ and the SM Higgs, the partial width of $h\to Z'Z'$ is simply proportional to $(g_{BL} \sin\theta)^2$ and given by a substitution of $m_\phi\to m_h$ in Eq.~(\ref{eq:phiZpZp}).
For small $g_{BL}$, the $Z'$ is long-lived and escapes the detector, but its contribution to the invisible Higgs decay width is tiny. For large $g_{BL}$, the $Z'$ is short-lived and decays inside the detector and does not contribute to the invisible Higgs decay width. We are not aware of any searches for $h\to Z'Z'\to 4 \ell$, which could provide a constraint for large gauge couplings.

\subsection{Additional \texorpdfstring{$Z'$}{Z'} production from \texorpdfstring{$\phi \to Z'Z'$}{f->Z'Z'}}
\label{sec:FASERimprovedZp}

When $m_\phi> 2m_{Z'}$, there exists an additional $Z'$ production mode through
$B\to K\phi$ or $K\to \pi\phi$ followed by the decay $\phi\to Z'Z'$.
The decays $b\to s\phi(\to Z'Z')$ and $s\to d\phi(\to Z'Z')$ can be described by the FCNC Lagrangian in Eq.~\eqref{eq:FCNCcoupling} and the $\phi Z' Z'$ coupling~\cite{Bird:2004ts,Altmannshofer:2009ma,Feng:2017vli}
\begin{eqnarray}
\mathcal{L}_{\phi Z'Z'}=Q'_\Phi g_{BL} m_{Z'} Z^{\prime \mu}Z'_\mu \phi\;.
\end{eqnarray}
We estimate the contribution from $B$ meson decays by considering the differential partial decay width for the inclusive decay $b\to s Z'Z'$
\begin{equation}
\begin{aligned}
{d\Gamma_{b\to sZ'Z'}\over dq^2}&=\frac{\sin^2\theta \,m_b^3m_{Z'}^2 Q_\Phi^{\prime2} g_{BL}^2}{256\pi^3v_0^2}
\left|\frac{3m_t^2 V_{ts}^\ast V_{tb}}{16\pi^2 v_0^2}\right|^2
\\
&\times
\frac{1}{(q^2-m_\phi^2)^2+\Gamma_\phi^2 m_\phi^2}
\Big(1-{q^2\over m_b^2}\Big)^2
\Big(1-{4m_{Z'}^2\over q^2}\Big)^{1/2}
\Big[2+{(q^2-2m_{Z'}^2)^2\over 4m_{Z'}^4}\Big]\;,
\end{aligned}
\end{equation}
where $q$ is the sum of the two $Z'$ bosons' four-momenta.
The total decay width $\Gamma_\phi$ depends on all parameters $m_\phi$, $\sin\theta$, $m_{Z'}$ and $g_{BL}$ in this case, as mentioned in the above subsection and thus the (differential) branching ratio does not simply scale with $g_{BL}^2$, but also depends on the total decay width $\Gamma_\phi$. Similarly, the expression for $K\to \pi  Z' Z'$ is
\begin{equation}
\begin{aligned}
\frac{d \Gamma_{K\to \pi Z'Z'}}{dq^2}
& = \frac{ \sin^2\theta \,m_s^2 m_{Z'}^2Q_\Phi'^2 g_{BL}^2 }{512\pi^3 m_K v_0^2}
 \left(\frac{m_K^2-m_\pi^2}{m_s-m_d}\right)^2   \left|\frac{3m_t^2 V_{td}^\ast V_{ts}}{16\pi^2 v_0^2}\right|^2
\\
& \times\frac{\lambda^{1/2}(1,m_\pi^2/m_K^2,q^2/m_K^2) |f_0^{K\pi}(q^2)|^2}{(q^2-m_\phi^2)^2 + m_\phi^2 \Gamma_\phi^2} \left(1-\frac{4m_{Z^\prime}^2}{q^2}\right)^{1/2} \left[2+\frac{(q^2-2m_{Z^\prime}^2)^2}{4m_{Z^\prime}^4}\right]
\end{aligned}
\end{equation}
and the $K\to \pi$ form factor is given by $f_0^{K\pi}(q^2)=0.96$~\cite{Boiarska:2019jym}.
We implemented the 3-body decay in FORESEE. To efficiently translate the result from the reference coupling $g_{BL}=1$ and to avoid generating events for each coupling, we rescale the result with the change in the total cross section. The 3-body decays turn out to be subdominant compared to 2-body decays for the benchmark scenarios which we considered and thus the resulting sensitivity regions of FASER/FASER2 remain unchanged.

\section{Conclusions}
\label{sec:Con}

We study the general $U(1)'$ models with non-universal lepton charges. A scalar singlet is introduced with non-zero $U(1)'$ charge to spontaneously break the $U(1)'$ gauge symmetry. This model with small masses and couplings can be tested in precision experiments such as FASER/FASER2, COHERENT and the long-baseline neutrino oscillation experiments DUNE and T2HK.
We explore the sensitivities of above precision experiments to the new gauge boson $Z'$ and the new CP-even scalar $\phi$. Our main findings are the following:
\begin{itemize}
\item
With non-universal lepton charges, distinctive reaches emerge in the regime of low $m_{Z'}$ and small gauge coupling for different $U(1)'$ charge setups. The COHERENT experiment and the future long-baseline neutrino oscillation experiments DUNE and T2HK also provide complementary sensitivities.
\item
For $m_\phi < 2m_{Z'}$, the search for the scalar $\phi$ at FASER/FASER2 is sensitive to the mixing angle between the scalar singlet and the SM Higgs, which is analogous to the dark Higgs search.
\item
In the case of $m_\phi > 2m_{Z'}$, the kinematically allowed decay $\phi\to Z' Z'$ changes the lifetime and decay rates of the scalar $\phi$. The probe reach highly depends the $Z'$ mass and the gauge coupling $g_{BL}$ as additional parameters.
With $\phi\to Z' Z'$ open, an additional production mode of $Z'$ appears which is however subdominant to other production modes.
\end{itemize}

\acknowledgments
We would like to thank Yi Cai and Yongchao Zhang for discussion and Maksym Ovchynnikov for providing the sensitivity curves for SHiP and HIKE-dump.
T.L. is supported by the National Natural Science Foundation of China (Grant No. 11975129, 12035008) and ``the Fundamental Research Funds for the Central Universities'', Nankai University (Grant No. 63196013). J.L. is supported by the National Natural Science Foundation of China (Grant No. 11905299, 12275368) and Guangdong Basic and Applied Basic Research Foundation (Grant No. 2020A1515011479).
T.F. and M.S. acknowledge support by the Australian Research Council Discovery Project DP200101470. This research includes computations using the computational cluster Katana supported by Research Technology Services at UNSW Sydney.

\appendix

\section{Texture zeroes and lepton mixing}
\label{sec:Mixing}
We first assume one non-zero charge for leptons, for instance $Q'_e=-3$, and then one has the following Yukawa matrix textures
\begin{align}
	y^D & = \begin{pmatrix}
			*  & 0 & 0 \\
			0  & * & * \\
			0  & * & * \\
		\end{pmatrix}\;,
	    &
		M & = \begin{pmatrix}
			0  & 0 & 0 \\
			0  & * & * \\
			0  & * & * \\
		\end{pmatrix}\;,\\
		y^{M}(Q'_\Phi=3) & = \begin{pmatrix}
			0  & * & * \\
			*  & 0 & 0 \\
			*  & 0 & 0 \\
		\end{pmatrix}\;,
		    &
			y^{M}(Q'_\Phi=6) & = \begin{pmatrix}
			*  & 0 & 0 \\
			0  & 0 & 0 \\
			0  & 0 & 0 \\
		\end{pmatrix}\;,
\end{align}
with the following ranks
\begin{eqnarray}
&&\mathrm{rank}[y^D]  = 3\;,~
\mathrm{rank}[M]  = 2\;,~
\mathrm{rank}[y^M(Q'_\Phi=3)]  = 2\;,~
\mathrm{rank}[y^{M}(Q'_\Phi=6)]  =1\;,\nonumber \\
&&\mathrm{rank}[M+y^M(Q'_\Phi=3)]  = 3\;,~
\mathrm{rank}[M+y^M(Q'_\Phi=6)]  = 3\;.
\end{eqnarray}
In this case the assignment of $Q'_\Phi=3$ leads to the neutrino mass matrix without zeros, but $Q'_\Phi=6$ allows two zero entries.
Next we take two non-zero charges, for instance $Q'_e+Q'_\mu=-3$ with $Q'_e\neq 0$ and $Q'_\mu\neq 0$. For the choice of $Q'_e=Q'_\mu=-3/2$, we can have the Yukawa matrices
\begin{align}
	y^D & = \begin{pmatrix}
			*  & * & 0 \\
			*  & * & 0 \\
			0  & 0 & * \\
		\end{pmatrix}\;,
	    &
		M & = \begin{pmatrix}
			0  & 0 & 0 \\
			0  & 0 & 0 \\
			0  & 0 & * \\
		\end{pmatrix}\;,\\
		y^{M}(Q'_\Phi=3) & = \begin{pmatrix}
			*  & * & 0 \\
			*  & * & 0 \\
			0  & 0 & 0 \\
		\end{pmatrix}\;,
		    &
			y^{M}(Q'_\Phi=3/2) & = \begin{pmatrix}
			0  & 0 & * \\
			0  & 0 & * \\
			*  & * & 0 \\
		\end{pmatrix}\;,
\end{align}
with the following ranks
\begin{eqnarray}
&&\mathrm{rank}[y^D]  = 3\;,~
\mathrm{rank}[M]  = 1\;,~
\mathrm{rank}[y^M(Q'_\Phi=3)]  = 2\;,~
\mathrm{rank}[y^{M}(Q'_\Phi=3/2)]  =2\;,\nonumber \\
&&\mathrm{rank}[M+y^M(Q'_\Phi=3)]  = 3\;,~
\mathrm{rank}[M+y^M(Q'_\Phi=3/2)]  = 2\;.
\end{eqnarray}
One can see that, without introducing additional scalars, the assignment of $Q'_\Phi=3/2$ cannot give rise to three massive light neutrinos in this case. The choice of $Q'_\Phi=3$ can generate a full rank mass matrix, but there are still two zeros in the neutrino mass matrix. By including two scalar fields $\Phi, \Phi'$ with $Q'_\Phi=3$ and $Q'_{\Phi'}=3/2$, one can generate a full rank mass matrix and remove the zeros in the neutrino mass matrix.
For the case of $Q'_e+Q'_\mu=-3$ and $Q'_e\neq Q'_\mu$, we have
\begin{align}
	y^D & = \begin{pmatrix}
			*  & 0 & 0 \\
			0  & * & 0 \\
			0  & 0 & * \\
		\end{pmatrix}\;,
	    &
		M & = \begin{pmatrix}
			0  & 0 & 0 \\
			0  & 0 & 0 \\
			0  & 0 & * \\
		\end{pmatrix}\;,
        &
		y^{M}(Q'_\Phi=3) & = \begin{pmatrix}
			0  & * & 0 \\
			*  & 0 & 0 \\
			0  & 0 & 0 \\
		\end{pmatrix}\;,
\end{align}
with the ranks as
\begin{eqnarray}
\mathrm{rank}[y^D]  = 3\;,~
\mathrm{rank}[M]  = 1\;,~
\mathrm{rank}[y^M(Q'_\Phi=3)]  = 2\;,~
\mathrm{rank}[M+y^M(Q'_\Phi=3)]  = 3\;.
\end{eqnarray}
It turns out that there are two zero entries in the neutrino mass matrix.
Other assignments of $Q'_\Phi$ give at most one non-zero entry in $y^M$ and thus there must be two zeros in neutrino mass matrix.
For the case with three non-zero charges such as $Q'_e=Q'_\mu\neq Q'_\tau$, one has the following matrices
\begin{align}
	y^D & = \begin{pmatrix}
			*  & * & 0 \\
			*  & * & 0 \\
			0  & 0 & * \\
		\end{pmatrix}\;,
	    &
		M & = \begin{pmatrix}
			0  & 0 & 0 \\
			0  & 0 & 0 \\
			0  & 0 & 0 \\
		\end{pmatrix}\;,\\
		y^{M}(Q'_\Phi=-(Q'_e+Q'_\tau)) & = \begin{pmatrix}
			0  & 0 & * \\
			0  & 0 & * \\
			*  & * & 0 \\
		\end{pmatrix}\;,
        &
		y^{M}(Q'_\Phi=-2Q'_e) & = \begin{pmatrix}
			*  & * & 0 \\
			*  & * & 0 \\
			0  & 0 & 0 \\
		\end{pmatrix}\;,
\end{align}
with the ranks as
\begin{eqnarray}
&&\mathrm{rank}[y^D]  = 3\;,~
\mathrm{rank}[M]  = 0\;,~
\mathrm{rank}[y^M(Q'_\Phi=-(Q'_e+Q'_\tau))]  = \mathrm{rank}[y^M(Q'_\Phi=-2Q'_e)] =2\;,\nonumber \\
&&\mathrm{rank}[M+y^M(Q'_\Phi=-(Q'_e+Q'_\tau))] = \mathrm{rank}[M+y^M(Q'_\Phi=-2Q'_e)] = 2\;.
\end{eqnarray}
We find that this case cannot generate three massive neutrinos and one needs additional scalar fields to give a full rank mass matrix.
For the most general case with $Q'_e\neq Q'_\mu \neq Q'_\tau$, such as $Q'_e=-3$ and $Q'_\mu=-Q'_\tau\neq 3$ in particular, we have the following Yukawa matrices
\begin{align}
	y^D & = \begin{pmatrix}
			*  & 0 & 0 \\
			0  & * & 0 \\
			0  & 0 & * \\
		\end{pmatrix}\;,
	    &
		M & = \begin{pmatrix}
			0  & 0 & 0 \\
			0  & 0 & * \\
			0  & * & 0 \\
		\end{pmatrix}\;,\\
		y^{M}(Q'_\Phi=-2Q'_e) & = \begin{pmatrix}
			*  & 0 & 0 \\
			0  & 0 & 0 \\
			0  & 0 & 0 \\
		\end{pmatrix}\;,
        &
		y^{M}(Q'_\Phi=-(Q'_e+Q'_\mu)) & = \begin{pmatrix}
			0  & * & 0 \\
			*  & 0 & 0 \\
			0  & 0 & 0 \\
		\end{pmatrix}\;,\\
		y^{M}(Q'_\Phi=-2Q'_\mu) & = \begin{pmatrix}
			0  & 0 & * \\
			0  & * & 0 \\
			*  & 0 & 0 \\
		\end{pmatrix}\;,
        &
		y^{M}(Q'_\Phi=2Q'_\mu)) & = \begin{pmatrix}
			0  & 0 & 0 \\
			0  & 0 & 0 \\
			0  & 0 & * \\
		\end{pmatrix}\;,
\end{align}

For the most general case with $Q'_e\neq Q'_\mu \neq Q'_\tau$, we have the Yukawa matrices
\begin{align}
	y^D & = \begin{pmatrix}
			*  & 0 & 0 \\
			0  & * & 0 \\
			0  & 0 & * \\
		\end{pmatrix}\;,
	    &
		M & = \begin{pmatrix}
			0  & 0 & 0 \\
			0  & 0 & 0 \\
			0  & 0 & 0 \\
		\end{pmatrix}\;,\\
		y^{M}(Q'_\Phi=-(Q'_e+Q'_\tau)) & = \begin{pmatrix}
			0  & 0 & * \\
			0  & 0 & 0 \\
			*  & 0 & 0 \\
		\end{pmatrix}\;,
        &
		y^{M}(Q'_\Phi=-2Q'_e) & = \begin{pmatrix}
			*  & 0 & 0 \\
			0  & 0 & 0 \\
			0  & 0 & 0 \\
		\end{pmatrix}\;,
\end{align}
with the ranks as
\begin{eqnarray}
&&\mathrm{rank}[y^D]  = 3\;,~
\mathrm{rank}[M]  = 0\;,~
\mathrm{rank}[y^M(Q'_\Phi=-(Q'_e+Q'_\tau))]  = 2\;,~\mathrm{rank}[y^M(Q'_\Phi=-2Q'_e)] =1\;,\nonumber \\
&&\mathrm{rank}[M+y^M(Q'_\Phi=-(Q'_e+Q'_\tau))] = 2\;,~\mathrm{rank}[M+y^M(Q'_\Phi=-2Q'_e)] = 1\;.
\end{eqnarray}
Again, the additional scalar fields are needed to give a full rank mass matrix.

It turns out that adding one scalar singlet can lead to three massive light neutrinos and no zeros in the neutrino mass matrix in the case with one non-zero lepton charge. In other non-universal models, additional scalar fields are needed to generate a full rank mass matrix and remove the zero entries in the neutrino mass matrix. Alternatively, we could generate the missing entries in the Dirac neutrino Yukawa matrix. The cases with one or two independent zeros were also discussed in the literature, where either one or two scalar singlets are needed. To summarize, it is quite arbitrary to include the scalar singlets breaking the non-universal gauge symmetry to give the allowed PMNS matrix.

\bibliographystyle{JHEP}
\bibliography{refs}

\end{document}